\documentclass[aps, prb,  twocolumn,superscriptaddress]{revtex4-1}
\usepackage{graphicx}
\usepackage{dcolumn}
\usepackage{color}
\usepackage{epsfig}
\usepackage{setspace}
\usepackage{amsmath,amssymb}
\usepackage{verbatim}
\usepackage{bibentry}
\usepackage{bbold}
\usepackage{soul}
\usepackage{hyperref}

\def\nnu{{\nonumber}}
\def\bk{{\mathbf{k}}}
\def\bK{{\mathbf{K}}}

\def\bR{{\mathbf{R}}}
\begin{document}
\title{ A typical medium cluster approach for multi-branch phonon localization}
\author{Wasim Raja Mondal}
\affiliation{Department of Physics and Astronomy, Middle Tennessee State University, Murfreesboro, TN 37131, USA}

\author{ Tom\ Berlijn}
\affiliation{Center for Nanophase Materials Sciences, Oak Ridge National Laboratory, Oak Ridge, Tennessee 37831, USA}

%\author{Yi\ Zhang}
%\affiliation{Department of Physics \& Astronomy, Louisiana State University, Baton Rouge, LA 70803, USA}

\author{N.\ S.\ Vidhyadhiraja} 
\affiliation{Jawaharlal Nehru Centre for Advanced Scientific Research, Bangalore 560 064, India.}

\author{Hanna Terletska} \email{hanna.teletska@mtsu.edu}
\affiliation{Department of Physics and Astronomy, Middle Tennessee State University}

\begin{abstract}
The phenomenon of Anderson localization in various disordered media has sustained significant interest over many decades. Specifically, the Anderson localization of phonons has been viewed as a potential mechanism for creating
fascinating thermal transport properties in materials. However, despite extensive work, the influence of the vector nature
of phonons on the Anderson localization transition has not been well explored. In order to achieve such an understanding, we extend a recently developed phonon dynamical cluster approximation (DCA) and its typical medium variant (TMDCA)
to investigate spectra and localization of multi-branch phonons
in the presence of pure mass disorder. We
validate the new formalism against several limiting cases and exact diagonalization results.
A comparison of results for the single-branch versus multi-branch case shows that the vector nature of the
phonons does not affect the Anderson transition of phonons significantly. The developed multi-branch
TMDCA formalism can be employed for studying phonon localization in real materials.
\end{abstract}

\date{\today}
\maketitle

\section{Introduction}
Anderson localization~\cite{PhysRev.109.1492,kramer2012localization,PhysRevLett.42.673,PhysRevLett.48.699} of phonons~\cite{PhysRevB.27.5592,10.1063/5.0073129} has emerged as an area of great interest recently, owing to its potential for fundamental research as well as practical applications in fields such as thermoelectrics~\cite{Tian2019,Leeacs}, phononics~\cite{PhysRevLett.117.025503} and beyond. For example, in thermoelectric materials, Anderson localization can serve as a mechanism to enhance thermal transport properties. Specifically, the ability to localize phonons can be used as a tool to enhance the figure of merit ($zT$) by reducing the contribution of phonons to thermal transport through phonon localization. For instance, the $zT$ value of $\text{C}_{60}/\text{Cu}_2\text{Se}$ nanocomposites has been found to increase by 20-30$\%$ compared to that of pure Cu$_2$Se, achieving a value of 1.4 at 773K~\cite{Tian2019,Leeacs}. This was achieved by reducing thermal conductivity through phonon scattering and increasing the Seebeck coefficient by carrier localization, which originates from the incorporation of C$_{60}$ nanoparticles. Very recently, Yang et. al.\cite{YANG2024} experimentally achieved ultralow thermal conductivity (0.19 W/mK) and a record average $zT$ (1.2 at room temperature) for Mg$_3$(Sb,Bi)$_2$ materials by utilizing phonon-localization. A coupled phonon-magnon localization is found to enhance function near ferroic glassy states\cite{doi:10.1126/sciadv.adn2840}. In the field of nanotechnology, phonon localization has been actively explored in reduced dimensions \cite{Ma_2020, PhysRevX.10.021050, PhysRevB.81.224208} in superlattice structures, where it has been effectively used to enhance their performance and capabilities \cite{doi:10.1126/sciadv.aat9460}. 
Despite the challenges in achieving phonon localization
due to the broad spectrum of heat-carrying phonons, there have been reports of its direct observation in several technologically important materials, as well as in artificially made elastic networks. For example, Manley $et$  $al.$\cite{Manley2014} observed ferroelectric phonon localization in relaxor ferroelectric PMN-30$\%$PT using neutron scattering. Howie $et$  $al.$~\cite{PhysRevLett.113.175501} observed phonon localization in a dense hydrogen-deuterium binary alloy. Faez $et$  $al.$\cite{PhysRevLett.103.155703} experimentally revealed the criticality of the Anderson localization transition for phonons. Phonon localization has also been observed in a random three-dimensional elastic network~\cite{Hu2008}. Luckyanova $et$  $al.$~\cite{doi:10.1126/sciadv.aat9460} observed phonon localization in GaAs/AlAs superlattices in the presence of randomly distributed ErAs nanodots at the interfaces. Very recently, Islam $et$  $al.$~\cite{Islam2022} found localization of optical phonon modes in boron nitride nanotubes.

All these experimental observations call for theoretical investigations of phonon localization in realistic models involving complex degrees of freedom, particularly considering the multi-branch vector nature of phonons. The direction of vibrations can have a significant impact on the Anderson transition. For example, the vector character of light may suppress Anderson localization of light, at least for point scatters\cite{PhysRevLett.112.023905}. However, such a prominent effect of the direction of vibrations on Anderson transition has not been found for the elastic wave considering elastic wave scattering\cite{PhysRevB.98.064206}.
The theory of elasticity is primarily concerned with continuous media and long-wavelength phenomena. In contrast, optical phonons involve discrete atomic displacements, which limits the direct applicability of elastic theory to these modes. Nevertheless, optical phonons are the most susceptible to Anderson localization, as discussed in Ref.~\cite{PhysRevB.96.014203}. Consequently, a comprehensive investigation of Anderson localization that incorporates the microscopic atomic details of phonons is essential for understanding phonon localization.

Various numerical methods, including continuum field theory~\cite{PhysRevB.27.5592}, diagrammatic technique~\cite{PhysRevB.31.5746}, exact diagonalization(ED)~\cite{PhysRev.154.802,PhysRevB.81.224208}, 
replica method\cite{PhysRevB.28.6358, PhysRevB.27.5592}, multifractal analysis(MFA)~\cite{PhysRevB.67.132203}, transfer matrix method~\cite{Pinski_2012} have been employed to investigate phonon localization, particularly in simplified model systems.
\textcolor{black}{
While real-space finite-size approaches provide unbiased benchmarks and controlled scaling, a practical constraint is the rapid growth in computational cost required to reach system sizes large enough to capture disorder effects accurately. For example, system sizes up to $18^3$     \cite{PhysRevB.81.224208}, $48^3$ ~\cite{PhysRevB.67.132203}, and $200^3$ \cite{PhysRevB.98.064206} are typically required, reflecting the exponential growth of the Hilbert space, the need for broad scaling windows and robust statistics, and the asymptotically large $L$ necessary to extract critical behavior via finite-size scaling.
%One of the main limitations of real-space finite-size methods is their rapidly increasing computational cost, which arises from the need to simulate very large system sizes to capture disorder effects accurately. 
%For example, ED~\cite{PhysRevB.81.224208}, MFA~\cite{PhysRevB.67.132203}, and finite-size scaling analyses~\cite{PhysRevB.98.064206} have typically required system sizes up to $18^3$, $48^3$, and $200^3$, respectively.
}

As an alternative class to real space finite-size approaches, effective medium theories with controllable approximations provide a self-consistent framework for treating disorder. Within this framework, the coherent potential approximation (CPA)~\cite{PhysRev.156.809,RevModPhys.46.465} serves as the canonical single site scheme, where a translationally invariant medium is determined self consistently from disorder averaged quantities. Building on CPA, cluster extensions including the dynamical cluster approximation (DCA)~\cite{PhysRevB.63.125102,PhysRevB.89.081107,PhysRevB.90.094208,PhysRevB.92.014209,PhysRevB.92.205111,PhysRevB.94.224208} embed a finite cluster of size $N_c$ in a self-consistently determined effective medium, which incorporates short-range nonlocal correlations and improves accuracy for both electronic and vibrational models. For lattice dynamics, the itinerant coherent potential approximation (ICPA)~\cite{PhysRevB.66.214206} extends CPA to include mass and force constant disorder and has been used to describe vibrational spectra in realistic binary alloys.

%As an alternative class to real-space finite-size approaches,
%To overcome such computational cost, various approximation schemes with controllable approximations have been developed based on effective medium theory. Coherent Potential Approximation (CPA)~\cite{PhysRev.156.809, RevModPhys.46.465}. and its cluster extensions, including the Dynamical Cluster Approximation (DCA) ~\cite{PhysRevB.63.125102, PhysRevB.89.081107,PhysRevB.90.094208,PhysRevB.92.014209,PhysRevB.92.205111,PhysRevB.94.224208,2021} have been successfully employed to investigate both disordered electronic and phononic systems. The itinerant coherent-potential approximation (ICPA~)~\cite{PhysRevB.66.214206} has been used to investigate the disorder effect on the vibrational spectra in realistic binary alloys. While these effective medium approaches can describe certain precursor phenomena associated with localization, they are unable to capture Anderson localization itself.

\textcolor{black}{Although CPA and DCA accurately reproduce disorder-averaged spectra, they do not provide a description of Anderson localization~\cite{PhysRevB.89.081107}. The typical-medium dynamical cluster approximation (TMDCA) addresses this by incorporating geometric averaging of the local density of states within the cluster self-consistency, so that the typical density of states (TDOS) serves as an order parameter that vanishes at localization~\cite{PhysRevB.89.081107}. In TMDCA, a finite cluster of size $N_c$ is embedded in a self-consistent effective medium; the approximation is controllable since $1/N_c$ acts as a small parameter, and increasing $N_c$ systematically adds nonlocal correlations, yielding convergence toward the exact result. This construction accesses thermodynamic-limit disorder averages without explicit finite-size scaling or prohibitively large real-space lattices. For the three-dimensional Anderson tight-binding model, TMDCA reproduces the critical disorder strength $W_c$ and the localization phase diagram for electron localization with box disorder in excellent agreement with transfer-matrix method (TMM) results using moderate clusters (e.g., $N_c=92$)~\cite{PhysRevB.89.081107}.  In practice, the Green’s function-based TMDCA offers computational
efficiency and flexibility that make it particularly well-suited for investigating complex disordered systems. These include systems with multiple orbitals, long-range disorder potentials, electron-electron interactions, and features often inaccessible to conventional finite-size approaches~\cite{app8122401, PhysRevB.94.224208,PhysRevB.89.081107, PhysRevB.90.094208, PhysRevB.92.014209,PhysRevB.94.235104,TAM2021168480, TERLETSKA2021168454, cryst11111282}.} 

\textcolor{black}{
While Anderson localization of electrons arises from interference of scattered electronic wave functions, the corresponding phenomenon for phonons occurs in the vibrational spectrum of disordered lattices. In such systems, disorder effects are most pronounced at high frequencies (short wavelengths), whereas low-frequency acoustic modes remain extended even under strong disorder. Numerical studies using finite-size methods, such as ED~\cite{PhysRevB.81.224208}, MFA~\cite{PhysRevB.67.132203}, and finite-size scaling ~\cite{PhysRevB.98.064206} on large system sizes, together with experimental observations~\cite{PhysRevLett.113.175501}, uch delocalization of long-wavelength phonons. These studies further demonstrate that the localization transition of high-frequency modes belongs to the same universality class as electronic Anderson localization~\cite{PhysRevB.81.224208,PhysRevB.67.132203,PhysRevB.98.064206}. 
Consequently, the effective medium TMDCA, which is well established for electronic systems, can be extended to study phononic systems~\cite{PhysRevB.96.014203}. In contrast to the very large lattices required for the finite-size studies, TMDCA reproduces the phonon localization phase diagram in quantitative agreement with transfer-matrix-method (TMM) results using modest clusters (e.g., $N_c=4^3$) while retaining computational efficiency~\cite{PhysRevB.96.014203}. TMDCA has been established to be a successful method for studying Anderson localization of phonons in all disorder limits~\cite{PhysRevB.96.014203}. However, previous TMDCA implementations for phononic systems with mass~\cite{PhysRevB.96.014203} and force-constant~\cite{PhysRevB.99.134203} disorder have been restricted to single-branch models, i.e., vibrations in a single polarization direction.
}

\textcolor{black}{Despite this progress, modeling phonon localization in realistic materials remains challenging.} An effective numerical method must construct material-specific force-constant models and solve the resulting lattice-dynamical problem. The absence of lattice translational symmetry in disordered solids and the breakdown of the Bloch theorem make this task difficult to solve straightforwardly. Moreover, force-constants calculations are computationally expensive compared to electronic case hopping calculations, and maintaining the force-constants sum rule can be challenging, especially in the presence of disorder. The inclusion of vibration directions adds another layer of complexity to the problem due to the correlations, where the vibration of an atom at site $R_l$ in the $\alpha$ direction is coupled with the vibration of an atom at site $R_{l^\prime}$ in the $\beta$ direction. 
Furthermore, accurately capturing Anderson localization in such systems is challenging, as it often requires the calculation of two-particle transport quantities or finite-size scaling analysis that can be computationally expensive, particularly for complex realistic models. ~\cite{PhysRevB.95.144208}. \textcolor{black}{ Nonetheless, first-principles-based workflows have recently begun to tackle phonon localization in specific materials systems.~\cite{PhysRevLett.118.145701,coatings12040422}}

%In this paper, to enable a more realistic modeling of phonon systems and their localization within the effective medium framework, 
%we have developed the multi-branch cluster DCA and its typical medium variant, TMDCA. The DCA and TMDCA, are two cluster-effective medium methods that were originally developed for electronic systems~\cite{PhysRevB.63.125102, PhysRevB.89.081107,PhysRevB.90.094208,PhysRevB.92.014209,PhysRevB.92.205111,PhysRevB.94.224208,2021}, and are used to study the disorder effects and localization, respectively.
%In the present work, building upon our previous work on phonon localization in a mass-disordered binary alloy using a single-direction vibration model~\cite{PhysRevB.96.014203}, 
%we have extended the TMDCA formalism to three directions of vibrations
%without taking into account force-constant disorder\cite{PhysRevB.99.134203}. 
\textcolor{black}{
In this work we extend TMDCA to treat phonons with three polarization directions, enabling more realistic modeling of vibrational spectra and localization within an effective-medium framework. This development builds on earlier TMDCA studies of phonon localization in single-branch models with mass disorder~\cite{PhysRevB.96.014203} and with force-constant disorder~\cite{PhysRevB.99.134203}. Here we formulate a multi-branch ( with three directions of vibrations) TMDCA for mass-disordered lattices.
We validate the developed formalism against exact-diagonalization benchmarks and appropriate limiting cases. Applying our developed multi-branch formalism, we investigate how the vector nature of vibrations affects the spectra and localization of phonons. After comparing the results of single-branch and multi-branch situations, we found that the Anderson transition of phonons is not significantly influenced by the vector nature of the phonons for the specific model considered in this work. Overall, the present study provides further understanding of the localization of vector phonons and brings the TMDCA one step closer to being an efficient numerical tool for studying Anderson localization of phonons in materials from first principles.   }

The paper is organized as follows. In section~\ref{sec: model and formalism}, we describe the model, the Green's function approach we utilize to solve our problem, and provide a detailed description of our multi-branch DCA and TMDCA cluster approaches. In section~\ref{sec: results}, we present our results and their discussion. We conclude in section~\ref{sec: conclusion}.

\section{Formalism and Model}
\label{sec: model and formalism}

%\subsection{Formalism}

\subsection*{ A. Hamiltonian}

We consider the lattice vibrational Hamiltonian in the harmonic approximation, given as
\begin{equation}
H=\sum_{\alpha, i, l}\frac{{\big(p^{\alpha}_{i}(l)}\big)^{2}}{2M_{i}(l)} + \frac {1}{2} \sum_{\alpha,\beta, l,l^\prime, i,j} \Phi^{\alpha\beta}_{ij}(l,l^\prime)u_{i}^{\alpha}(l)u_{j}^{\beta}(l\prime),
\label{eq:ham1}
\end{equation}
here $p_{i}^{\alpha}(l)$ and $u^{\alpha}_i(l)$ represent, respectively, the momentum and the displacement (from the equilibrium) of the $i^{\text{th}}$ atom with mass $M_i(l)$ in a unit cell occupying the lattice site $R_l$ vibrating along ``branch" $\alpha=(x,y,z)$ directions. 
The index $i$ represents the atom's number in the unit cell with $N_{cell}$ being the maximum number of atoms in the unit cell; the index $l$ denotes the atom position on a lattice. 
The force-constant $\Phi^{\alpha\beta}_{ij}(l,l^\prime)$ couples the $i^{\text{th}}$ atom at the lattice site $R_l$ vibrating along the $\alpha$ direction with the $j^{\text{th}}$ atom at the lattice site $R_{l^\prime}$ vibrating along the $\beta$ direction. Note, that the mass $M_i(l)$ in Eq.~\ref{eq:ham1} can vary randomly from site to site and, in general, can be modeled by different types of disorder distributions\cite{PhysRevB.96.014203}.

\subsection*{ B. Green's function formalism}
Applying the equation of motion method to the Hamiltonian of Eq.~\ref{eq:ham1}, we obtain the equation for the phonon Green's function, $D^{\alpha\beta}_{ij}(l,l^\prime,\omega)$, in the following form

\begin{align}
M_i(l) \omega^2 D^{\alpha\beta}_{ij}(l,l^\prime,\omega) &= \delta_{\alpha\beta}\delta_{ll^\prime}\delta_{ij}\mathbb{1} \nnu \\
&+\sum_{\gamma, l^{\prime\prime},j^\prime} {\Phi^{\alpha \gamma}_{i j^\prime}(l, l^{\prime\prime})} D^{\gamma \beta}_{j^\prime j}(l^{\prime\prime}, l^\prime,\omega),
\label{eq:eom1}
\end{align}
here, $D^{\alpha\beta}_{ij}(l, l^\prime,\omega)$ is the retarded displacement-displacement phonon Green's function\cite{RevModPhys.46.465}, defined as
\begin{equation}
iD_{ij}^{\alpha\beta}(l,l^\prime,\omega)=\langle\langle\, u_{i}^{\alpha}(l,\omega);u_{j}^{\beta}(l^\prime,\omega)\, \rangle\rangle. 
\label{eq:gfapp}
\end{equation}
$\langle\langle ... \rangle\rangle$ stands for the expectation value and time ordering between two displacement ($u^{\alpha}_i(l)$) and ($u^{\beta}_j(l)$) operators.

Eq.~\ref{eq:eom1} can be rewritten in terms of the Dyson's equation using the ``dressed'' $D_{ij}^{\alpha\beta}(l,l^\prime,\omega)$ and ``bare'' non-disordered phonon Green's functions $D_{ij}^{\alpha\beta (0)}(l,l^\prime,\omega)$ as follows
\begin{equation}
{\bigg(D_{ij}^{\alpha\beta}(l,l^\prime,\omega)\bigg)}^{-1} = \left(D_{ij}^{\alpha\beta (0)}(l,l^\prime,\omega)\right)^{-1} - \omega^2{V}_{ij}^{\alpha\beta}(l,l^\prime)\, ,
\label{eq:dyson}
\end{equation}
where the  mass-disorder potential is given as 
\begin{equation}
{V}_{ij}^{\alpha\beta}(l,l^\prime) = \left(1 - \frac{M_{i}(l)}{M_0}
\right) \delta_{\alpha\beta} \delta_{ij}\delta_{ll^\prime}\,.
\label{eq:dispot}
\end{equation}
Note, that the mass-disorder potential is local and identical in all directions  ($x,y,z$), i.e., it is completely diagonal in both the branch and spatial bases.

In the zero mass-disorder limit, the mass does not vary from site to site, corresponding to $M_i(l)=M_0$, and hence the system becomes translationally invariant. In such a case, Eq.~\ref{eq:eom1}, when expressed in momentum 
$\bk$-space, reduces to
\begin{equation}
{{D}}_{ij}^{\alpha\beta (0)}(\bk,\omega) = \left[\omega^2\mathbb{1} - {F}_{ij}^{\alpha\beta}(\bk)\right]^{-1},
\label{eq:D0}
\end{equation}
here, ${F}^{\alpha\beta}_{ij}(\bk)$ is related to the force-constants $\Phi^{\alpha\beta}_{ij}(l,l^\prime)$ as
\begin{equation}
 {F}^{\alpha\beta}_{ij}(\bk) =\sum_{l^\prime} \frac{\Phi^{\alpha\beta}_{ij}(l,l^\prime)}{M_0} e^{i\bk\cdot(\bR_l - \bR_{l^\prime})}\,.
\label{eq:Fmat1}
\end{equation}
In our current implementation, we consider the host mass $M_0$ to be unity and account for one atom per unit cell considering just the nearest neighbors ($R_{l^\prime}=R_{l\pm1}$). Therefore, from here on, we will omit the atom indices $i,j$, and the host-mass index $M_0$ in our discussions.

\subsection*{ C. Dynamical Cluster Approximation for multi-branch phonons}
\label{subsec:DCA-details}

%To solve Eq.~\ref{eq:dyson}, we have extended our previously developed single-branch phonon DCA method~\cite{PhysRevB.96.014203} to the given multi-branch case. 

In this section, we provide a detailed description of our extension of the Dynamical Cluster Approximation (DCA) to the multi-branch phonon case.
The DCA, as outlined in~\cite{PhysRevB.63.125102,RevModPhys.77.1027}, serves as a cluster extension of the CPA method, a commonly used single-site effective medium approach employed to investigate the impact of disorder in models and materials~\cite{PhysRev.156.809, RevModPhys.46.465}. In the CPA, the original lattice problem is mapped into an impurity problem embedded in an effective medium, which is determined self-consistently. In the absence of disorder, this is equivalent to the dynamical mean-field theory for strongly correlated systems (DMFT) \cite{RevModPhys.68.13}, a well-established and powerful tool for handling many-body strong correlation effects in clean systems. The DCA, as a cluster extension of the CPA, systematically incorporates non-local spatial correlations that are absent in the CPA. These correlations have been identified as crucial for accurately describing the effects of disorder~\cite{TERLETSKA2021168454}.

The DCA is a momentum-space cluster extension of the CPA, in which the original 
lattice with $N$ sites is mapped onto a periodic cluster of size $N_c = L_c^d$ ($L_c$ is the linear dimension of the cluster, $d$ is the
the dimensionality of the system) 
embedded into a self-consistently determined effective medium.
In reciprocal space, this is equivalent to the division of
the Brillouin zone of the underlying lattice into $N_c$ cells
of size $(2\pi/L_c)$, centered at the reciprocal cluster vectors
$\bf K$.~\cite{PhysRevB.63.125102,RevModPhys.77.1027} The lattice momenta within a given cell are denoted by $\mathbf{\tilde{k}}$. In the self-consistency loop, the DCA cluster Green's function is obtained from the $\bf k$-dependent lattice quantities via coarse-graining over the momenta $\mathbf{\tilde{k}}$, with $\bf k=K+\tilde{k}$. After disorder-averaging, the translational symmetry is restored, 
%and the clusters are subject to periodic boundary conditions,
allowing one to use the usual lattice Fourier transform to obtain the disorder-averaged cluster quantities either from reciprocal space into real space or vice versa using the Fourier transform. The DCA algorithms recovers local CPA for $N_c = 1$, and becomes exact as $N_c \rightarrow \infty$. Therefore, with increasing
the cluster size $N_c$, the DCA systematically interpolates between
the single-site and the exact result while remaining in the
thermodynamic limit. We refer the reader to Ref.~\cite{PhysRevB.63.125102,RevModPhys.77.1027} for further details.

In the following, we describe the self-consistency numerical procedure used in our multi-branch extension of the DCA formalism. In the DCA, the effective medium is characterized by a non-local hybridization function ${\Gamma}(\mathbf K,\omega)$, which is constructed from the calculation of the disorder-averaged cluster DCA Green's function. The analysis is performed iteratively until the cluster disorder-averaged Green's function becomes the same as the coarse-grained lattice Green's function within a certain numerical accuracy. In this work, we extend the single-branch DCA phonon algorithm~\cite{PhysRevB.96.014203} to the multi-branch phonon system. The resulting DCA numerical self-consistently iterative procedure is described as follows:

\noindent
1. We start with an initial guess of the hybridization ${\Gamma}_{\text{old}}^{\alpha\beta}(\mathbf{K},\omega)$.  
One can start with an initial guess of the effective medium hybridization function either by setting ${\Gamma}_{\text{old}}^{\alpha\beta}(\mathbf{K},\omega)=0$ or using the coarse-grained non-disordered Green's function
\begin{equation}
{{\Gamma}_{\text{old}}^{\alpha\beta} (\mathbf{K}, \omega)} = \omega^2 \mathbb{1} - \frac{N_c}{N}\sum_{\tilde{\bf k}}  {F}^{\alpha\beta}(\mathbf{K}+\mathbf{\tilde{k}}) - \Big \lbrack {{\bar{D}}^{\alpha\beta (0)}(\mathbf{K},\omega )} \Big \rbrack ^{-1},
\label{eq:gammamb}
\end{equation}
here ${{D}}^{\alpha\beta (0)}(\mathbf{K},\omega )=\frac{N_c}{N} \sum_{\tilde{ \bf k}} {{D}}^{\alpha\beta (0)}(\mathbf{K+\tilde{k}},\omega )$ is the coarse-grained bare Green's function of Eq.~\ref{eq:D0} (a bar denotes the coarse-grained quantities).

\noindent
2. To set up the cluster problem, we then compute the cluster-excluded Green's function
${\mathcal{D}}^{\alpha\beta}(\mathbf{K}, \omega)$ of the effective medium
as 
\begin{equation}
{{\mathcal{D}}^{\alpha\beta}(\mathbf{K}, \omega)} = \Big \lbrack {\omega^2 \mathbb{1} -  {F}^{\alpha\beta}(\mathbf{K}) - {{\Gamma}_{\text{old}}^{\alpha\beta}(\mathbf{K}, \omega)}\Big\rbrack}^{-1} .
\label{eq:GcKintmb}
\end{equation}
For each cluster vector $\bK$, ${\mathcal{D}}$ is $N_b \times N_b$ matrix; $N_b=3$ corresponds to three branch directions of the vibration. 

\noindent
3. Next, to solve the cluster problem (with the disorder distributed randomly at cluster sites $L,L'$), we first Fourier transform  Eq.~\ref{eq:GcKintmb} to the real-space form of the cluster
\begin{equation}
{{\mathcal{D}}^{\alpha\beta}(L,L^\prime, \omega)} = \sum_{\mathbf{K}} {{\mathcal{D}}^{\alpha\beta}(\mathbf{K}, \omega)} \exp^{i\mathbf{K}(\mathbf{R}_L - \mathbf{R}_{L^\prime})}.
\label{eq:GcKrsmb}
\end{equation}
%here $L,L'$ indicate specific elements of the real-space cluster matrices.
\noindent
4. We then construct the cluster Green's function for a given disorder configuration $V$:
\begin{align}
&{{D}^{\alpha\beta}(V,L,L^\prime,\omega)} \nonumber\\
& = \sqrt{\mathbb{1} - V(L,L)} \Bigg \lbrack  \Big\lbrack  { {\mathcal{D}}^{-1}(\omega)}  - \omega^2 \hat{V}  \Big \rbrack^{\alpha\beta}_{LL^\prime} \Bigg \rbrack^{-1} \sqrt{ \mathbb{1}- V(L^\prime,L^\prime)},
\label{eq:multdw}
\end{align}
%
%\begin{align}
%&{\hat{D}^{c,\alpha\beta}(V,L,L^\prime,\omega)} \nonumber\\
%& = \sqrt{1 - V_L} \Bigg \lbrack  \Big\lbrack \lbrack {\hat {\mathcal{D}}(\omega)} \rbrack^{-1} - \omega^2 \hat{V}  \Big \rbrack^{\alpha\beta}_{LL^\prime} \Bigg \rbrack^{-1} \sqrt{1 - V_{L^\prime}},
%\label{eq:multdw}
%\end{align}
here $\mathcal{D}$, ${D}$  are $N_bN_c \times N_b N_c$ matrices in the real space of the cluster. In our analysis, we consider a binary alloy disorder described by the following  distribution function
\begin{equation}
P(V (L,L))= c_A \delta(V(L,L) - V_A) + c_B \delta(V(L,L) - V_B),  
\end{equation}
here $c_A$ is the concentration of $A$ atoms and $c_B = (1-c_A)$ is the concentration of $B$ atoms at a given cluster site $L$, and $V_A=-V_B$. 

\noindent
5. Next, we perform the averaging over disorder and  calculate the disorder-averaged cluster Green's function
\begin{equation}
{ D^{\alpha\beta}_{ave}(L,L^\prime,\omega)}  =  \bigg\langle  { D^{\alpha\beta}(V,L,L^\prime,\omega)}  \bigg\rangle ,
\label{eq:cluavgnmb}
\end{equation}
where $<...>$ indicates an averaging over disorder configurations.

\noindent
6. The disorder-averaged cluster Green's function of Eq.~\ref{eq:cluavgnmb} is then Fourier transformed from the real space to the momentum space ${D^{\alpha\beta}_{ave}(\bf K,\omega)}$ and is used to calculate the coarse-grained lattice Green's function 
\begin{align}
{\bar{D}}^{\alpha\beta}(\mathbf{K},\omega) & = \frac{N_c}{N} \sum_{\tilde k} \Big\lbrack \Big ( {D}^{\alpha\beta}_{ave}(\mathbf{K},\omega) \Big)^{-1} \nonumber \\
& + {\Gamma}_{\text{old}}^{\alpha\beta}(\mathbf{K},\omega) - 
{F}^{\alpha\beta}(\mathbf{K}+\mathbf{\tilde{k}}) + {\bar F}^{\alpha\beta}(\mathbf{K}) \Big \rbrack^{-1} 
\label{eq:latg}
\end{align}
%Here $\hat{\bar{D}}^{\alpha\beta}(\mathbf{K},\omega)$ is $N_b \times N_b$ matrix for each $\bK$.
The DCA self-consistency condition requires that the disorder-averaged cluster Green's function equals to the coarse-grained lattice Green's functions, once the convergence is reached  
\begin{equation}
    {\bar{D}}^{\alpha\beta}(\mathbf{K},\omega)={D}^{\alpha\beta}_{ave}(\mathbf{K},\omega)
\end{equation}

\noindent
7. 
A new hybridization function is then obtained using the updated coarse-grained lattice Green's function 
\begin{align}
{\Gamma_{\rm new}^{\alpha\beta} (\bK,\omega)} & = {\Gamma_{\rm old}^{\alpha\beta} (\bK,\omega)}+ \nonumber \\
& \xi \left ( {\bar{D}}^{\alpha\beta}(\mathbf{K},\omega) -\left({ D^{{ \alpha\beta}}_{ave} (\bK,\omega)}\right)^{-1}\right),
\label{eq:gammanewmb}
\end{align}
here $0\leq\xi\leq 1$ is a linear mixing parameter used for improving the numerical convergence. We repeat the iteration procedure until the hybridization function converges to the desired numerical accuracy with ${  \Gamma_{\rm old}^{\alpha\beta} (\bK,\omega)}={\Gamma_{\rm new}^{\alpha\beta} (\bK,\omega)} $. 
%When this happens, the cluster Green's function and the coarse-grained lattice Green's function are also converged within the computational error, ${\hat {\bar{D}}^{\scriptscriptstyle \alpha\beta} (\mathbf K,\omega)}={\hat {D}^{\scriptscriptstyle c, \alpha\beta}_{ave} (\mathbf K,\omega)}$. 

In our analysis to determine the effect of disorder on spectral properties, we use the obtained DCA disorder-averaged Green's function ${D}^{\scriptscriptstyle{\alpha\beta}}_{ave}(\bK,\omega)$ to calculate the arithmetic average density of states (ADOS) for a given phonon branch $\alpha$ as
\begin{equation}
{\rm ADOS}^{\alpha\alpha}(\omega^2)=-\frac{2\omega}{N_c\pi}{\rm Im}
\sum_\bK \Big \lbrack {D}^{\alpha\alpha}_{ave}(\bK,\omega^2)\, \Big\rbrack .
\label{eq:dca-dos}
\end{equation}

However, as shown in Refs.~\cite{PhysRevB.96.014203,TERLETSKA2021168454}, the DCA-produced ADOS cannot distinguish between localized and extended states. Being the arithmetically averaged quantity, it always favors the extended state and therefore is not critical at the Anderson transition. To identify the localized state, one then has to resort to the typical medium analysis~\cite{PhysRevB.89.081107,Dobrosavljevi} which we describe in the following section.

\subsection*{D. Typical medium dynamical cluster approximation for phonons}

To study the Anderson localization of phonons, one must adopt the typical medium approach in the analysis. It has been demonstrated~\cite{Dobrosavljevi, PhysRevB.96.014203, TERLETSKA2021168454} that the typical density of states (TDOS) serves as a suitable order parameter for detecting Anderson-localized states. TDOS vanishes for localized states and remains finite for extended states~\cite{Dobrosavljevi}. In the typical medium approach, the TDOS is approximated using geometric averaging over disorder configurations. Refs.~\cite{Dobrosavljevi, PhysRevB.96.014203, TERLETSKA2021168454} show that such geometrically averaged TDOS continuously diminishes as disorder strength approaches the critical point, making it an effective one-particle order parameter for detecting Anderson localization transition.

To incorporate such a typical medium approach into our cluster self-consistency loop, we replace the linearly averaged cluster Green's function $D^{\alpha\beta}_{ave}(L,L^\prime,\omega)$ of Eq.~\ref{eq:cluavgnmb} with the typical cluster Green's function $D^{\alpha\beta}_{typ}(L,L^\prime,\omega)$
obtained from the geometrically averaged density of states. For this purpose, we adopt the same ansatz as in the single-branch case~\cite{PhysRevB.96.014203} for computing the TDOS ($\rho_{typ}$), namely
\begin{equation}
\rho_{typ} (\mathbf{K},\omega) = e^{\frac{1}{N_c}\sum_L <\ln \rho_{LL}(\omega)> } \Bigg\langle \frac{\rho(\mathbf{K},\omega)}{\frac{1}{N_c}\sum_{L} \rho_{LL}(\omega) }  \Bigg\rangle    
\end{equation}
and extend this to the multi-branch case similarly to the multi-branch case for the electron systems~\cite{PhysRevB.92.205111} as: 

\begin{widetext}
\begin{equation}
{\rho^{\alpha\beta}_{typ}(\mathbf{K},\omega)}  =
\begin{pmatrix}
e^{\frac{1}{N_c}\sum_L <\ln \rho^{\alpha\alpha}_{LL}(\omega)> } \Bigg\langle \frac{\rho^{\alpha\alpha}(\mathbf{K},\omega)}{\frac{1}{N_c}\sum_{L} \rho^{\alpha\alpha}_{LL}(\omega) }  \Bigg\rangle & \ldots  & e^{\frac{1}{N_c}\sum_L <\ln |\rho^{\alpha\beta}_{LL}(\omega)|> } \Bigg\langle \frac{\rho^{\alpha\beta}(\mathbf{K},\omega)}{\frac{1}{N_c}\sum_{L} \rho^{\alpha\beta}_{LL}(\omega) }  \Bigg\rangle\\
\vdots & & \vdots\\
e^{\frac{1}{N_c}\sum_L <\ln |\rho^{\beta\alpha}_{LL}(\omega) |>} \Bigg\langle \frac{\rho^{\beta\alpha}(\mathbf{K},\omega)}{\frac{1}{N_c}\sum_{L} \rho^{\beta\alpha}_{LL}(\omega) }  \Bigg\rangle  &\ldots & e^{\frac{1}{N_c}\sum_L <\ln \rho^{\beta\beta}_{LL}(\omega)> } \Bigg\langle \frac{\rho^{\beta\beta}(\mathbf{K},\omega)}{\frac{1}{N_c}\sum_{L} \rho^{\beta\beta}_{LL}(\omega) }  \Bigg\rangle  \\
\end{pmatrix},
\label{eq:mult_anstz}
\end{equation}
\end{widetext}
where
\begin{align}
\rho^{\alpha\beta}_{LL}(\omega) &=-\frac {2 \omega}{\pi}{\rm Im} \,D^{\alpha\beta}(L,L,\omega)  \\
\rho^{\alpha\beta}(\bK,\omega) & = -\frac {2 \omega} {\pi}  {\rm Im}\, D^{\alpha\beta}(\bK,\omega) 
\end{align}
are the site local and momentum-dependent spectral functions, respectively, computed from the cluster Green's function of Eq.~\ref{eq:multdw}.

From Eq.~\ref{eq:mult_anstz}, the disorder-averaged typical cluster Green's
function is obtained using the Hilbert transform as
\begin{equation}
D_{\scriptscriptstyle{\rm typ}}^{\alpha\beta}(\mathbf K,\omega) = {\mathcal P} \int d\omega^\prime \frac {{\rho^{\alpha\beta}_{\rm typ} (\bK,\omega^\prime)}} {\omega^2-\omega^{\prime 2}}
-i \frac {\pi}{2\omega} {{\rho^{\alpha\beta}_{\rm typ} (\mathbf{K},\omega}})\,,
\label{eq:ht}
\end{equation}
The rest of the implementation of the multi-branch TMDCA algorithm is very similar to that of the multi-branch DCA of the previous subsection $C$. The only difference is that instead of solving for the algebraically averaged cluster Green's function $D^{\alpha\beta}_{ave}$, one calculates the typical cluster Green's function $D^{\alpha\beta}_{\rm typ}$ 
using Eq.~\ref{eq:ht}. Once the convergence is reached in the self-consistency loop, we calculate the local TDOS 
\begin{equation}
{\rm TDOS}^{\alpha\alpha}(\omega^2)=-\frac{2\omega}{N_c\pi}{\rm Im}
\sum_\bK \Big \lbrack D_{\scriptscriptstyle{\rm typ}}^{\alpha\alpha}(\bK,\omega) \Big\rbrack \,.
\label{eq:def-tdos}
\end{equation}
to analyze the localization of the phonons (see Sec.~\ref{sec: results}).

\subsection{Multi-branch force constants model}
To simulate the impact of the branch coupling on the localization of phonon, in the following, we first describe the multi-branch force constant model $\Phi ^{\alpha \beta}(l,l')$ that we use in our analysis. We consider the force constants between atoms vibrating in three different directions, namely $(x, y, z)$. For instance, $\Phi^{\alpha\beta}(l,l^\prime)$ represents the force constants between atom A at lattice site $R_l$, which vibrates along the $\alpha (x,y,z)$ direction, and atom B, which vibrates along the $\beta(x,y,z)$ direction. Consequently, the force-constant matrices comprise both diagonal components $\Phi^{\alpha\alpha}(l,l^\prime)$, associated with atoms vibrating along the same directions, and off-diagonal components $\Phi^{\alpha\beta}$, associated with atoms vibrating in different directions.

\begin{table}[h!]
\begin{tabular}{|c|c|}
\hline
Interaction & Value \\
\hline
$\Phi^{\alpha\alpha}(\Vec{R}_l, \Vec{R}_l)$ & -6 \\
\hline
$\Phi^{\alpha\alpha}(\Vec{R}_l \pm \Vec{a}_\alpha , \Vec{R}_{l})$ & $1+\varepsilon$
\\
\hline
$\Phi^{\alpha\alpha}(\Vec{R}_l \pm \Vec{a}_\beta , \Vec{R}_l)$ & $1 - \frac{\varepsilon}{2}$ \\
\hline
$\Phi^{\alpha\beta}(\Vec{R}_l, \Vec{R}_l)$ & $-6\times \tau$ 
\\
\hline
$\Phi^{\alpha\beta}(\Vec{R}_l \pm \Vec{a}_\gamma , \Vec{R}_l)$ & $\tau$ \\
\hline
\end{tabular}
\caption{The force constant model used in our study. $\Phi^{\alpha\beta}(\Vec{R_{l}}, \Vec{R_{l^\prime}})$ describes the coupling between the atom at the lattice-site $\Vec{R}_{l}$ vibrating along $\alpha$ direction and the atom at the lattice-site $\Vec{R}_{l^\prime}$ vibrating along $\beta$ direction, where $\alpha, \beta, \gamma \in \{x,y,z \}, \alpha \neq \beta$ , \text{and} $\Vec{a}_x, \Vec{a}_y$ and $\Vec{a}_z$ are the primitive cell cubic lattice vectors.}
\label{table:force_descr}
\end{table}

Table-\ref{table:force_descr} defines the specific model that we consider in our work. The top three rows of the table, parameterized by $\varepsilon$, 
describes components of the force constant that are diagonal in the branch space, but off-diagonal in the lattice-space. The bottom two rows describes the components that are off-diagonal in the branch basis, which are parametrized by $\tau$. 
The parameters $\varepsilon$ and $\tau$ are introduced in such a way that the sum rule is satisfied, i.e., $\sum_l \Phi^{\alpha\beta}(l,l^\prime)=0$. 

Notice, that our model recovers the following limited cases. For $\tau=0$ and $\varepsilon\ne 0$, the model reduces to the trivial case of the system with three decoupled anisotropic branches. When $\tau\ne 0$ and $\varepsilon= 0$, the model also reduces effectively to three decoupled models, each of which rescaled by the eigenvalues of the $3\times3$ onsite matrix $\Phi^{\alpha\beta}(l,l) $: $1-\tau, 1-\tau,1+2\tau$. Importantly, in this case the disorder-induced localization properties of each of these three decoupled models are identical to those of the single-branch model because the eigenvectors of a model are not affected by rescaling the model with an overall constant.  Therefore, to investigate a model that is non-equivalent to the single branch case we need to consider $\tau\ne 0$ and $\varepsilon\ne 0$.

\section{Results and discussions}
\label{sec: results}

%In our DCA and TMDCA calculations, we have consider $1000$ disorder realizations for $N_c=1$ and $100$ disorder realization for the cluster size $N_c=4^3$ on a simple cubic lattice keeping broadening factor $\eta=10^{-3}$. To test the cluster-size convergence of our results, we have also utilized cluster size $N_c=5^3$ and verified that our results converge for the clusters beyond $N_c=4^3$. 

\subsection*{ A. Benchmarking of the DCA and ED methods}

Since the exact results for the three-directional lattice vibrational model have not been reported in the literature, we have first performed calculations using the exact diagonalization (ED) method. In ED calculations, we adopt the large-size disordered supercells. Within the supercell, the disorder is randomly distributed with the impurity distributions periodically repeating beyond the supercell boundaries. 
%and beyond the supercell boundaries, the impurity distributions periodically repeat. 
In ED analysis, we derive the force constant matrices of 100 supercells each with 60 impurities and roughly 400 sites on average for each set of the model parameters. The dynamical matrix of each supercell is then computed and diagonalized on a $10 \times 10 \times 10$ supercell momentum space grid. In our DCA and TMDCA calculations, we consider $1000$ disorder realizations for $N_c=1$ and $100$ disorder realization for the cluster size $N_c=4^3$ on a simple cubic lattice keeping broadening factor $\eta=10^{-3}$. To test the cluster-size convergence of our results, we have also utilized cluster size $N_c=5^3$ and verified that our results converge for the clusters beyond $N_c=4^3$

\begin{figure}[!t]
\includegraphics[width=0.5\textwidth]{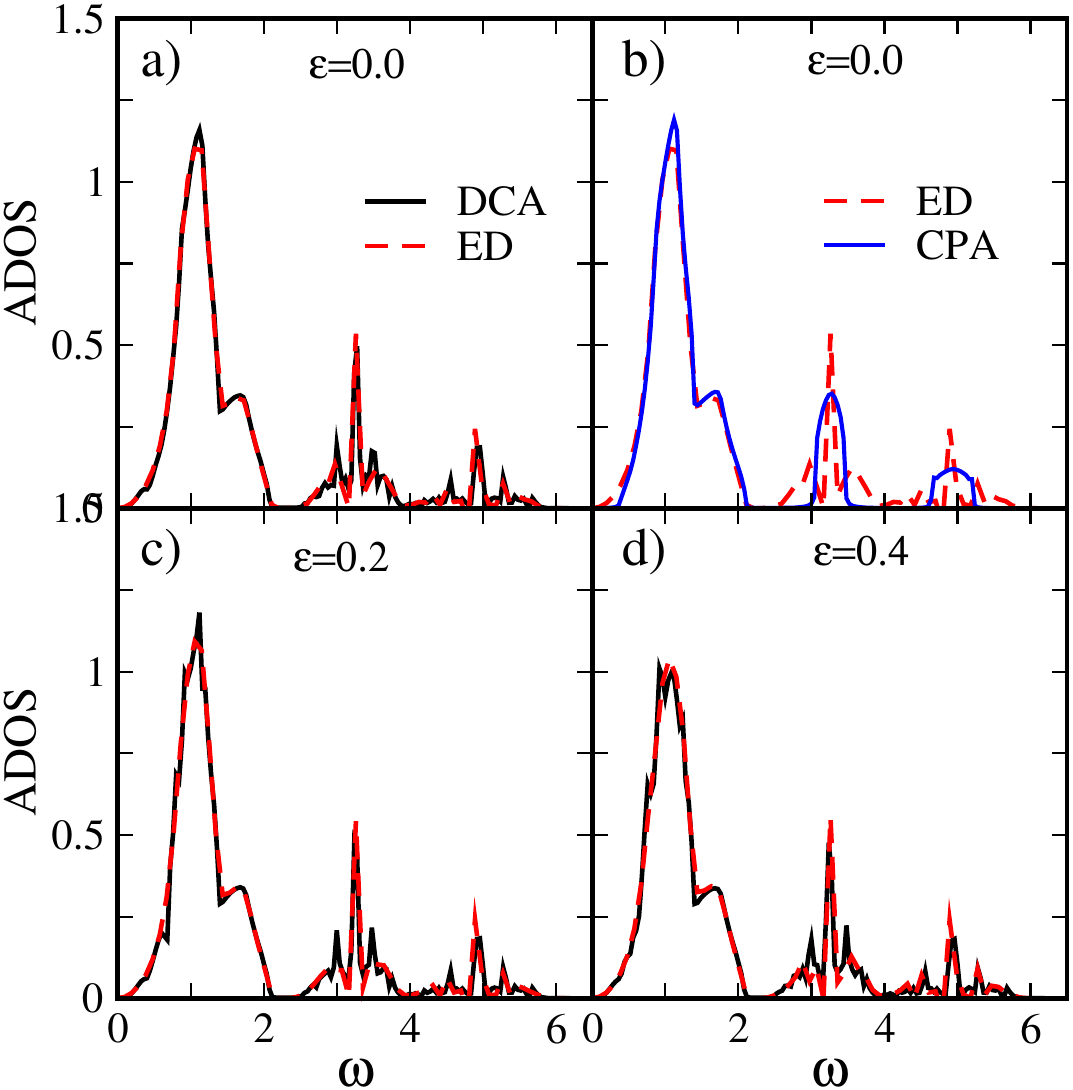}
\caption{{Parameters: $\tau=0.3, c_A=0.2,V_A=0.9$}. 
A comparison of the ADOS obtained by the developed multi-branch DCA, single site CPA and the exact diagonalization method (ED).
Panel a) shows the ADOS obtained from the multi-branch DCA using a cluster size of $N_c=4^3$ and the ED results for $\varepsilon=0$. Panel b) compares the single-site CPA results with ED results for $\varepsilon=0$. Note that, unlike the DCA, the CPA fails to reproduce the fine-structure features of the high-frequency impurity mode of the ADOS as found in exact calculations. Panels c) and d) show the multi-branch DCA results for the case where the atomic vibrations at lattice site $R_l$ in the direction $\alpha$ are coupled with atomic vibrations around $R_{l'}$ in the direction $\beta$ by the tuning parameter $\varepsilon$ of $0.2$ and $0.4$, respectively. It is observed that the host mode height changes with increasing $\varepsilon$, and the multi-branch DCA results are in strong agreement with the ED results. 
}
\label{fig:Fig1-2-3-eps-dep-benchmarks}
\end{figure}

We start the discussion of our results by first comparing our developed multi-branch cluster DCA method with the ED results for a binary isotopic alloy system in three dimensions. This benchmarking analysis aims to evaluate the accuracy and effectiveness of our developed multi-branch cluster DCA approach in capturing the system's behavior.
For this, in Fig.~\ref{fig:Fig1-2-3-eps-dep-benchmarks}, we compare the average density of states (ADOS) calculated for an impurity concentration of $c_A=0.2$ and a disorder potential of $V=0.9$, for a fixed value of $\tau=0.3$ and varying values of $\varepsilon$. First, on panel Fig.~\ref{fig:Fig1-2-3-eps-dep-benchmarks}-a), we consider the case of $\varepsilon=0.0$, where we compare the DCA and ED results, and observe a perfect agreement between two methods. For the finite values of $\tau$, we anticipate the emergence of two impurity modes in the higher frequency regions. Specifically, we expect a doubly degenerate impurity band at the lower end of the high-frequency region and a single impurity band at the higher end, corresponding to the eigenvalues of the $3\times3$ onsite matrix $\Phi^{\alpha\beta}(l,l) $: $1-\tau, 1-\tau,1+2\tau$ as explained in section ~\ref{sec: model and formalism}. As shown in Fig.~\ref{fig:Fig1-2-3-eps-dep-benchmarks}, the two impurity modes around frequency regions $2<\omega<4$ and $4<\omega<6$ are well captured by our DCA method and are in perfect agreement with the ED results. 

To evaluate the effects of non-local spatial correlations, Fig.~\ref{fig:Fig1-2-3-eps-dep-benchmarks}-b) compares the ADOS obtained with the local CPA method, corresponding to the 
 $N_c=1$ DCA case, against the exact ED data. Although the CPA method approximately reproduces the overall shape of the exact ADOS, it fails to capture the fine-structure features of the impurity modes, especially in the high-frequency region. This discrepancy is anticipated and well-understood due to the single-site nature of the CPA method, which cannot capture the non-local spatial correlations present in the system, leading to significant structures in the ADOS. These results underscore the importance of non-local spatial correlations, absent in the local CPA method, but accounted for in the finite cluster DCA.

Next, we proceed to compare our multi-branch DCA results against ED data for a finite value of $\varepsilon$. This corresponds to a scenario where atomic vibrations at site $R_{l}$ in the $\alpha$ directions are coupled with atomic vibrations around $R_{l^\prime}$ in the $\beta$ directions by the tuning parameter $\varepsilon$. Fig.~\ref{fig:Fig1-2-3-eps-dep-benchmarks}-c) and Fig.~\ref{fig:Fig1-2-3-eps-dep-benchmarks}-d) show the results for $\varepsilon=0.2$ and $\varepsilon=0.4$, respectively. While there is no significant difference in the high-frequency impurity modes, it is observed that the height of the host mode decreases with increasing $\varepsilon$ values. Our multi-branch DCA method successfully captures all the fine features of the ADOS, as evidenced by the remarkable agreement with ED data for a binary alloy.
As presented in Appendix A, we have also explored other disorder distributions, such as uniform distributions, and found excellent agreement between the multi-branch DCA results and ED results. These findings demonstrate the efficiency of our developed multi-branch DCA scheme in computing the average phonon density of states for complex systems.

It is worth noting that we have verified the convergence of our results with respect to the cluster size $N_c$. We found that the results obtained for $N_c=4^3$ and $N_c=5^3$ are practically identical, which establishes that our data converges very quickly to the exact thermodynamic limit. This highlights the efficiency of the DCA $N_c=4^3$ cluster in capturing the essential physics of complex systems while keeping computational costs reasonable.

\subsection*{B. $\varepsilon=0$ and $\tau\neq 0$: branch-decoupled case
%B. Branch-decoupled case: $\varepsilon=0$ and $\tau\neq 0$
}

\begin{figure}[!t]
\includegraphics[width=0.49\textwidth]{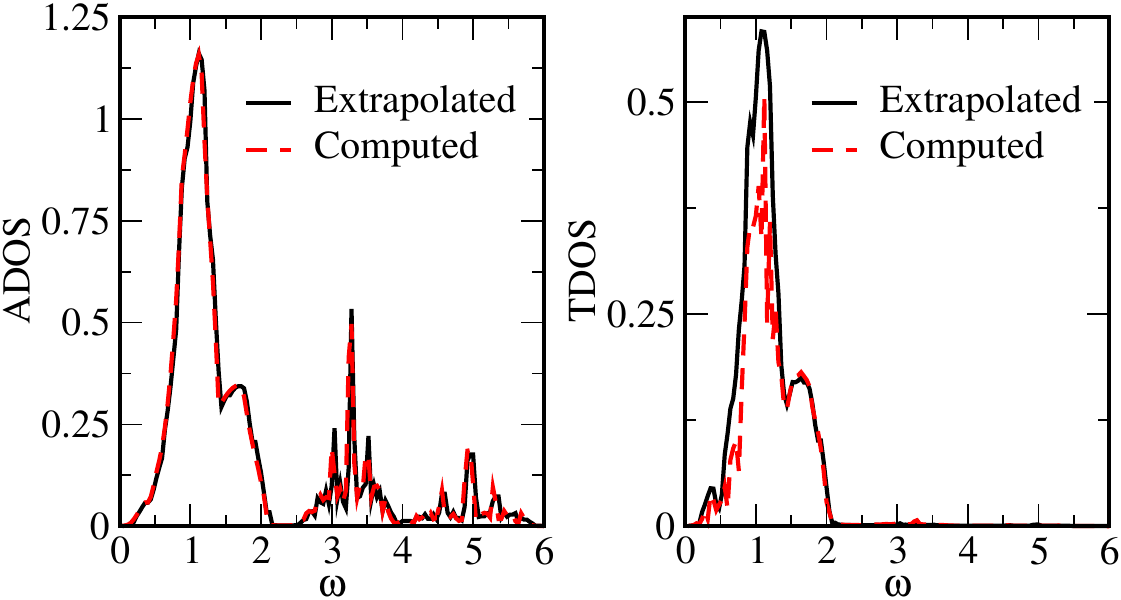}
\caption{Parameters: $\tau=0.3, \varepsilon=0.0, c_A=0.2, V_A=0.9$. Left panel: comparison of the ADOS obtained from Eq.~\ref{eq:dos_branch_relation} and from the multi-branch DCA formalism. Right panel: comparison of the TDOS obtained from Eq.~\ref{eq:tdos_branch_relation} and the multi-branch TMDCA formalism.}
\label{fig:ADOS_single_mult}
\end{figure}

As discussed in Section~\ref{sec: model and formalism}, our multi-branch model simplifies to three decoupled single-branch models when $\varepsilon=0$ and $\tau \neq 0$. Under these conditions, the total average density of states (ADOS) of the multi-branch model, can be obtained from the average density of states of a single branch, ados($\omega$) as follows (see Appendix B for the details): 
\begin{equation}
\text{ADOS}(\omega)= \sum_m\text{ados}(\frac{\omega}{\sqrt{\epsilon_m}}),
\label{eq:dos_branch_relation}
\end{equation}
where $\varepsilon_m$ are the eigenvalues of the $3\times3$ onsite matrix $\Phi^{\alpha\beta}(l,l) $, equal to $1-\tau, 1-\tau,1+2\tau$, respectively. 

To validate these results, Fig.~\ref{fig:ADOS_single_mult} (left panel) presents the ADOS obtained via DCA (computed) and the ADOS extrapolated from the single-branch density of states using Eq.~\ref{eq:dos_branch_relation} for the case where $\varepsilon=0$ and $\tau=0.3$. As observed in the figure, there is an excellent agreement between the results.

Naively, one might expect that a similar extrapolation formula exists for the typical medium case, wherein the total multi-branch TDOS($\omega$) is related to the typical density of states, tdos($\omega$), of an individual branch in a manner similar to that observed for the average case (see Appendix B for further details)
\begin{equation}
\text{TDOS}(\omega)= \sum_m
\text{tdos}(\frac{\omega}{\sqrt{\epsilon_m}})  
\label{eq:tdos_branch_relation}
\end{equation}

However, as seen from Fig.~\ref{fig:ADOS_single_mult} (right panel), the TDOS($\omega$) extrapolated from the tdos($\omega$) of a single branch, using Eq.~\ref{eq:tdos_branch_relation} only approximately agrees with the TDOS computed directly from the multi-branch model using the TMDCA.   
In Appendix B, we detail why Eq.~\ref{eq:tdos_branch_relation} is not universally valid.

\subsection*{C. $ \varepsilon \neq 0$ and $\tau\neq 0$: inter-branch coupling case
%The multibranch case that is inequivalent to the single branch case: $ \varepsilon \neq 0$ and $\tau\neq 0$
}
\label{sec:interbranch}

\begin{figure}[!t]
\includegraphics[width=0.49\textwidth]{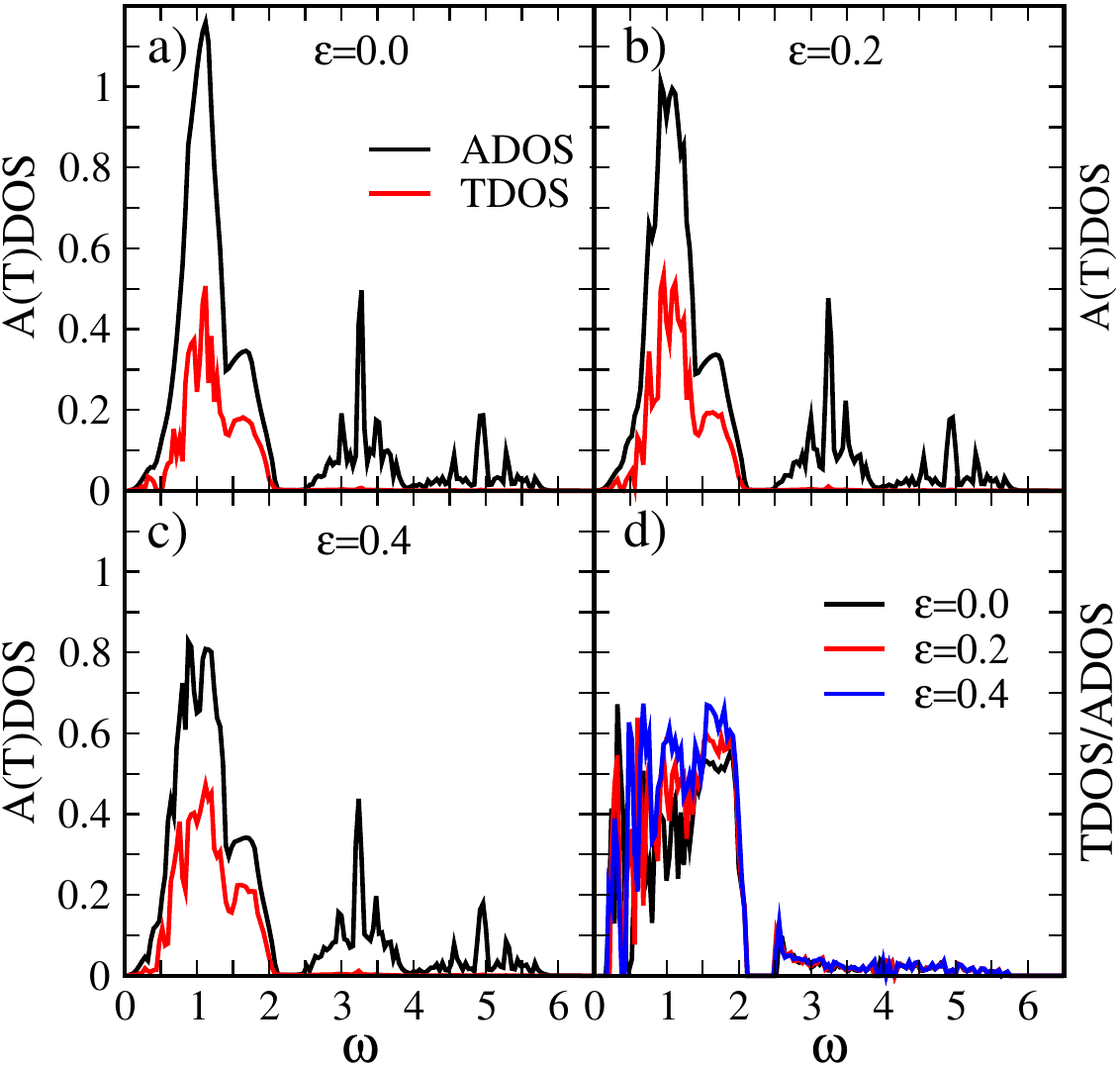}
\caption{Parameters: $\tau=0.3, c_A=0.2, V_A=0.9, N_c=4^3.$ Panels a), b), and c) show the evolution of the average ADOS and the typical TDOS for increasing values of $\varepsilon=0,0.2,0.4$ at at large disorder ($V_A=0.9$) and small impurity concentration $c_A=0.2$. Panel d) shows the ratio of TDOS/ADOS for all values of $\varepsilon=0,0.2,0.4$.}
%at at large disorder ($V=0.9$) and small impurity concentration $c=0.2$ 
\label{fig:epsilon_localized}
\end{figure}

In this section, we investigate the localization of phonons in our mass-disordered multi-branch model in the case where the inter-branch coupling occurs and the system is not equivalent to the single-branch model. This corresponds to the case with $\varepsilon\ne0$ and $\tau\ne0$.

In  Fig.~\ref{fig:epsilon_localized}, to demonstrate the impact of inter-branch coupling on phonon localization, we plot the ADOS and TDOS for increasing values of $\varepsilon=0.0, 0.2, 0.4$ at a fixed disorder potential $V_A=0.9$ and small impurity concentration of $c_A=0.2$. Beginning with the branch-decoupled case ($\varepsilon=0.0$), as shown in Fig.~\ref{fig:epsilon_localized}-a), we observe strong localization for high-frequency phonon modes within the frequency range of $\omega = 2.5$ to $\omega = 6.0$. This is evident from the TDOS approaching zero while the ADOS remains finite. The results for finite $\varepsilon=0.2$ and $\varepsilon=0.4$ are shown in Fig.~\ref{fig:epsilon_localized}-b) and Fig.~\ref{fig:epsilon_localized}-c). The data indicate a negligible effect of $\varepsilon$ on the localized phonons, although subtle changes are observed in the host modes. Specifically, the host mode becomes slightly broader and more delocalized with increasing $\varepsilon$. However, high-frequency phonons remain localized even as $\varepsilon$ increases. This is clearly illustrated in Fig.~\ref{fig:epsilon_localized}-d), where the ratio of TDOS/ADOS remains nearly constant across the range of considered $\varepsilon$ values ($0$ to $0.4$).

\begin{figure}[!t]
\includegraphics[width=0.5\textwidth]{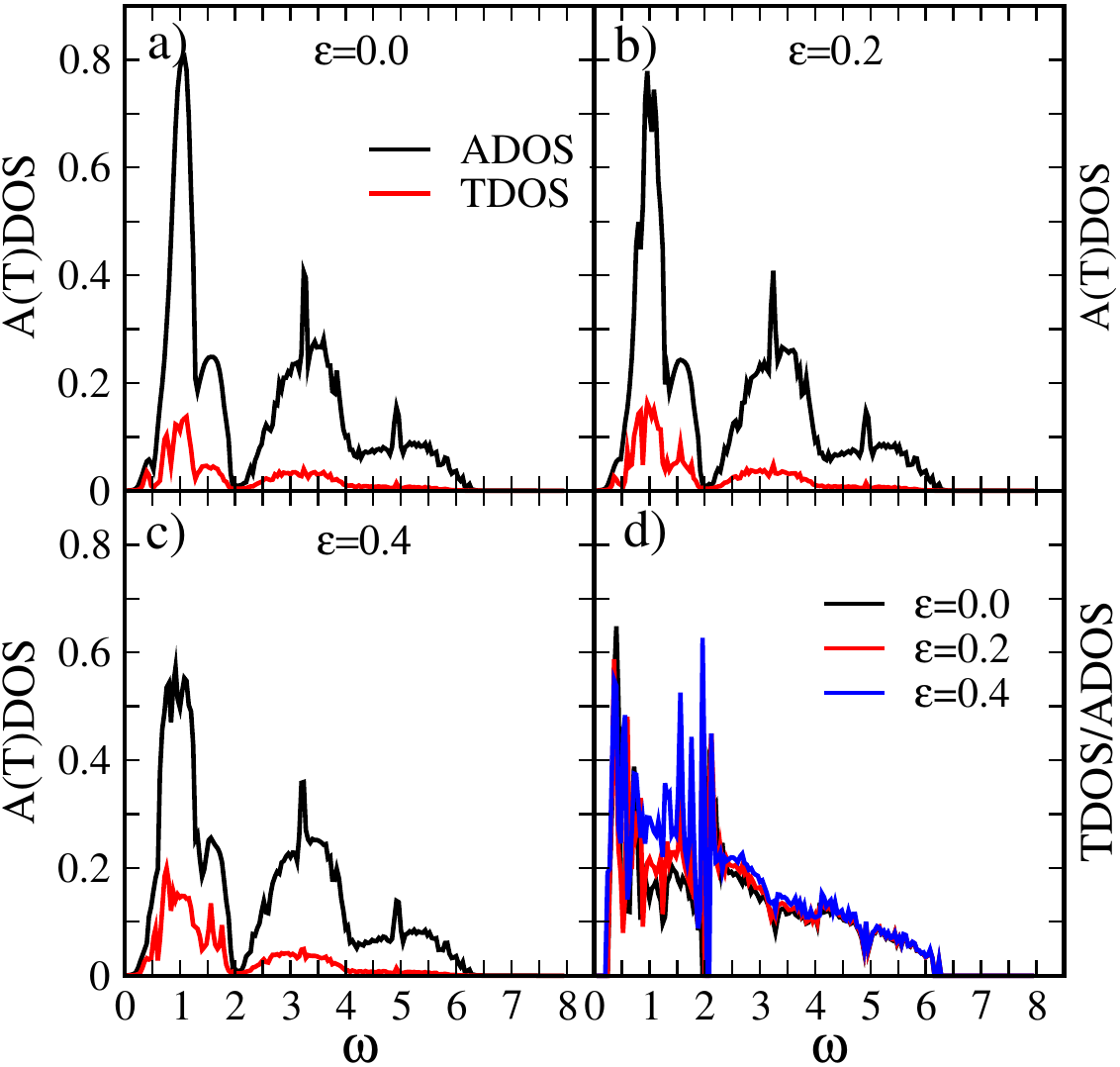}
\caption{Parameters: $\tau=0.3, c_A=0.5, V_A=0.9, N_c=4^3$. Panels a), b), and c) show the evolution of the average ADOS and the typical TDOS for increasing values of  $\varepsilon=0,0.2,0.4$ at large disorder ($V_A=0.9$) and moderate impurity concentration ($c=0.5$). d) the TDOS/ADOS ratio at $\varepsilon=0.0, 0.2, 0.4$. 
 }
\label{fig:epsilon_delocalized}
\end{figure}

Next, we consider an opposite scenario where we start with the impurity modes being delocalized, which occurs at higher impurity concentration at $c_A=0.5$ and $V_A=0.9, \varepsilon=0.0, 0.2, 0.4$ . In this case, as seen from Fig.~\ref{fig:epsilon_delocalized}-a) the impurity phonon modes are delocalized around $\omega=2$ to $4.5$ as depicted by a finite TDOS within this region. Our goal is to understand how $\varepsilon$ affects these delocalized phonons. As seen from Fig.~\ref{fig:epsilon_delocalized}-b) and Fig.~\ref{fig:epsilon_delocalized}-c), the delocalized phonons remain delocalized even when $\varepsilon$ is increased from $0$ to $0.4$, although there are subtle changes in the mid-frequency region of the spectrum. We confirm this observation by examining the ratio of TDOS/ADOS, which remains almost unchanged as shown in Fig.~\ref{fig:epsilon_delocalized}-d). 
\begin{comment}
{\color{black} Such behavior of states remaining delocalized is similar to what is observed in multi-orbital Anderson localization in electron systems, where mixing localized states with extended states keeps the system extended, consistent with Mott's insight about the mobility edge~\cite{Mott_1987}.}
\end{comment}

\begin{figure}[!t]
\includegraphics[width=0.5\textwidth]{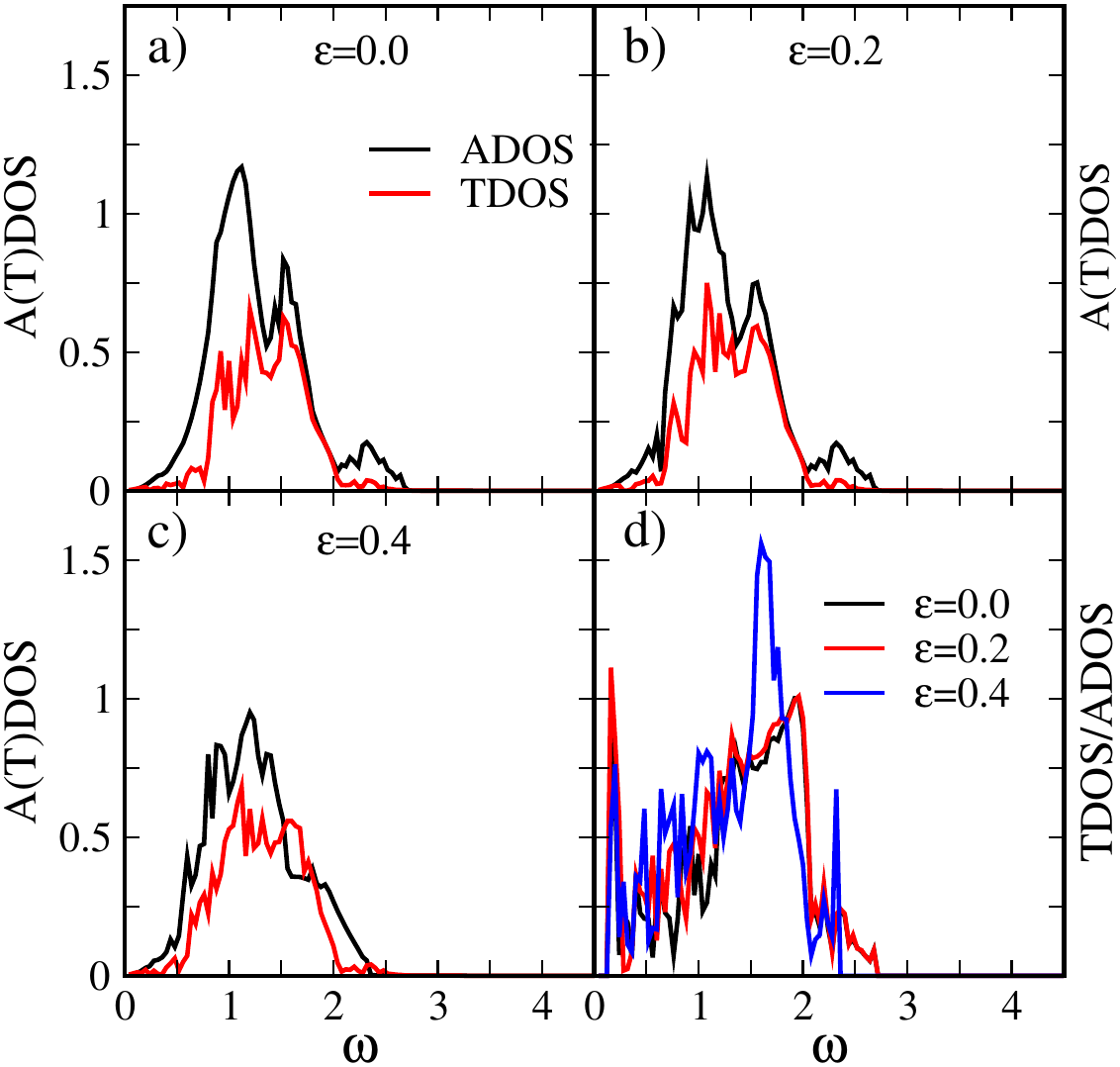}
% old Fig5.png
\caption{Parameters: $\tau=0.3, c_A=0.2, V_A=0.5, N_c=4^3$. 
Panels (a), (b), and (c) show the evaluation of ADOS) and TDOS and with increasing $\varepsilon$ values from 0 to 0.2 to 0.4 at intermediate disorder strength ($V_A=0.5$). (d) the TDOS/ADOS ratio at $\varepsilon=0.0, 0.2, 0.4$
}
\label{fig:epsilon_lowV}
\end{figure}

To further illustrate this effect, in Fig.~\ref{fig:epsilon_lowV}, we examine a case with a highly delocalized spectrum which occurs at low disorder potential $V_A=0.5$. For such case, we observe that ADOS and TDOS are both finite across the entire frequency range for increasing values of $\varepsilon$, with the delocalized phonon spectrum being concentrated in the mid-frequency region for all values $\varepsilon$.
Notably, while the TDOS shape shifts with increasing $\varepsilon$, the phonon modes remain delocalized. This is further confirmed by the near-constant ratio of TDOS/ADOS in Fig.~\ref{fig:epsilon_lowV}-d), indicating minimal impact of $\varepsilon$ on phonon localization. Therefore, a low disorder potential favors highly delocalized phonon modes unaffected by $\varepsilon$ changes.

\subsection*{D. 
%Effect of $ \varepsilon \neq 0$ and $\tau\neq 0$ on mobility edge }
Mobility edge analysis for box disorder}

\begin{figure}[!h]
%\vspace{-0.485in}
\includegraphics[width=0.5\textwidth]
{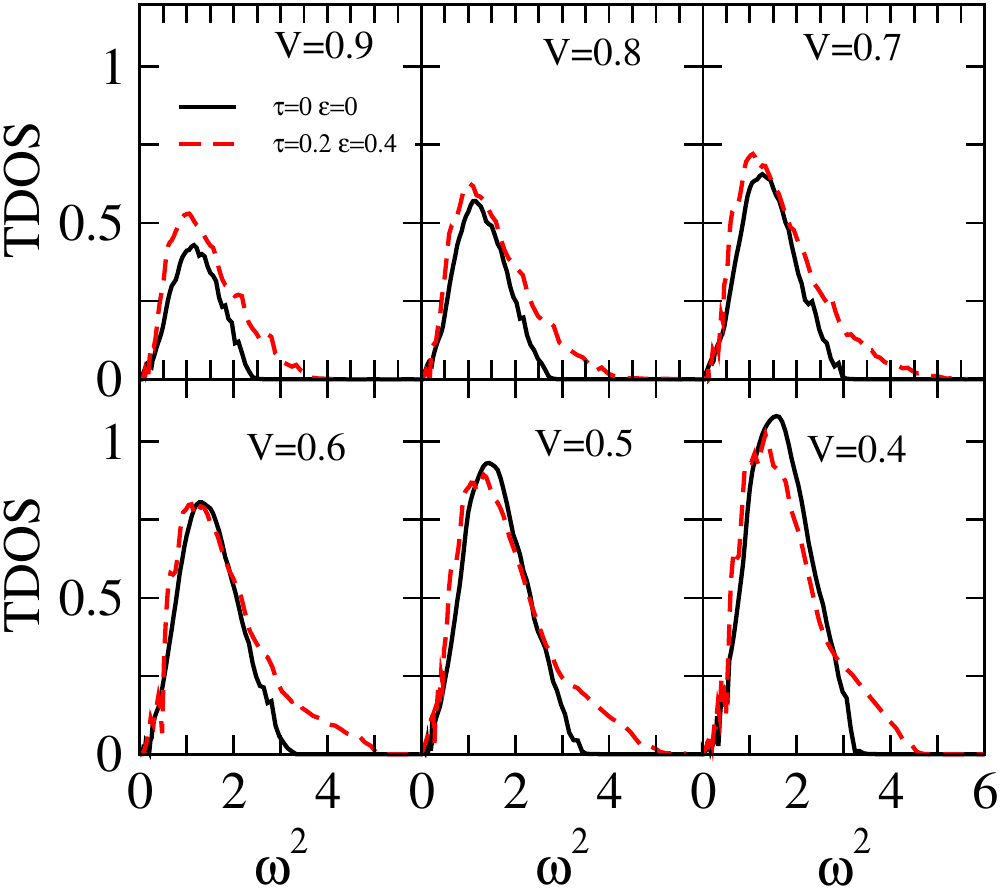}
\caption{Parameters: N$_c=4^3$, $\tau=0, 0.2$ and $\varepsilon=0, 0.4$. The evolution of the TDOS as a function of the square of the frequency $\omega^2$ at various disorder strengths $V=0.4, 0.5, 0.6, 0.7, 0.8, 0.9$ chosen from a box distribution. }
\label{fig:tdosbox}
%\vspace{-0.35in}
\end{figure}

\begin{figure}[!h]
%\vspace{-0.485in}
\includegraphics[width=0.45\textwidth]
{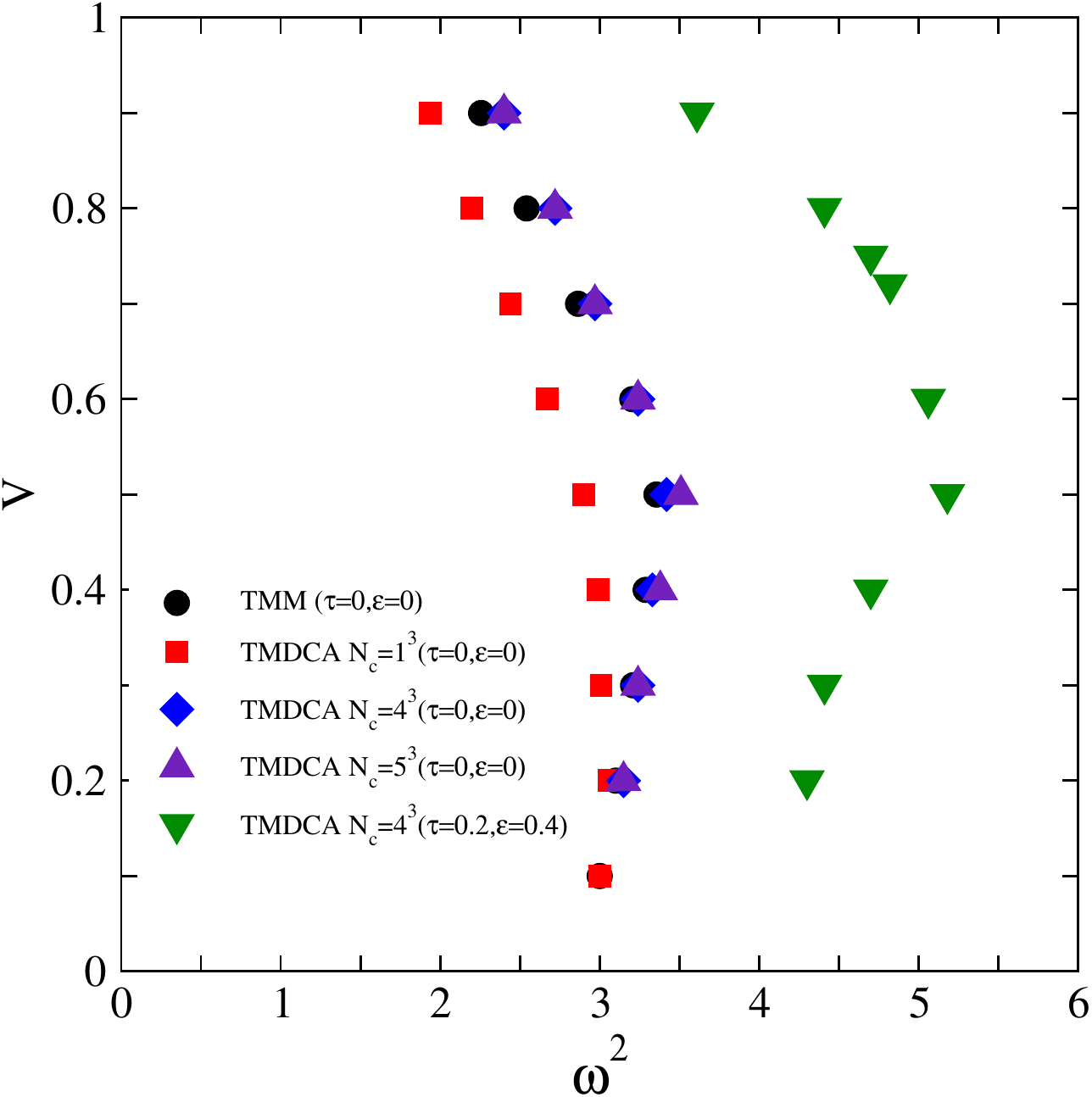}
%{figure/Fig1_3.png}
\caption{The mobility edge trajectories of the phonon spectrum in three dimensions for a box distribution.}
\label{fig:mobilityedge}
%\vspace{-0.35in}
\end{figure}

Finally, we investigate the influence of coupling parameters on the \textcolor{black} {mobility edge~\cite{PhysRev.109.1492, PhysRevLett.42.673,phystod50,PhysRevLett.134.053601}}, a critical quantity that separates localized and delocalized states in three-dimensional disordered systems. To this end, we consider a box disorder distribution defined as $P_V(V_l) = \Theta(V - | V_l |)/ 2V$, where $V_l = (1 - M_l/M_0)$, and $0 \leq V \leq 1$, where $l$ is the lattice site.

We identify the mobility edge as the high-frequency band edge of TDOS($\omega$), which marks the transition from extended to localized vibrational states. Fig.~\ref{fig:tdosbox} presents the TDOS for both the decoupled case ($\tau=0, \varepsilon=0$) and a finite coupling case ($\tau=0.2, \varepsilon=0.4$) for various strengths of the disorder $V$. As seen in the figure, the TDOS is increasingly suppressed with higher disorder $V$ for both coupling regimes. Notably, in the presence of finite branch coupling, the TDOS broadens, and the high-frequency band edge shifts to higher energies, indicating a modification of the mobility edge.

Using the TDOS data from Fig.~\ref{fig:tdosbox}, we can now construct the mobility edge trajectories on a $V$ vs. $\omega^2$ phase diagram. Our results are shown in Fig.~\ref{fig:mobilityedge}. We first benchmark our TMDCA mobility edge results against existing data obtained using the exact    {\color{black}Transfer Matrix Method (TMM)~\cite{Pinski_2012,warwick57512}} for the uncoupled case ($\tau=0, \varepsilon=0$). We find excellent agreement between the TMM data (black circles) and our TMDCA results for cluster size (blue diamonds). To examine cluster size dependence, we also include results for N$_c=5^3$, which closely overlap with the N$_c=4^3$ data, indicating convergence of the mobility edge trajectories. In contrast, the single-site results (N$_c=1$), based on the local Typical Medium Theory approximation\cite{Dobrosavljevi}, show significant deviations from the exact TMM trajectory and fail to reproduce the re-entrant behavior of the mobility edge. As observed in the TMM data, the mobility edge initially moves outward with increasing disorder strength in the range $V=0.1$ to $0.5$, before reversing direction and moving inward as $V$ increases further. This re-entrant behavior occurs because, at low disorder, localized states form outside the band edges, analogous to deep trap states. As disorder $V$ increases to intermediate values, these states hybridize both with each other and with nearby extended states, leading to partial delocalization and pushing the mobility edge outward. Beyond  $V=0.5$ further increase in disorder strength localizes the edge states, causing the mobility edge to move inward. The failure of the single-site N$_c=1$ approximation to capture this behavior arises from its lack of inter-site spatial correlations, which are essential for describing such hybridization effects. In contrast, the TMDCA, even with a modest cluster size of N$_c=4^3$, accurately captures the non-local physics responsible for the re-entrant behavior and the correct mobility edge trajectory. This highlights the necessity of finite cluster analysis for accurate capturing spatial correlations critical to phonon localization in disordered systems.

We now turn our discussion to the coupled scenario with $\tau=0.2, \varepsilon=0.4$. As shown in Fig.~\ref{fig:mobilityedge}, finite inter-branch coupling shifts the high-frequency edge of the TDOS toward higher values for any given disorder strength $V$. For example, at $V=0.4$, the right band edge of the TDOS shifts from $\omega^2=3.33$ to $\omega^2=4.70$, and at $V=0.9$, it shifts from $\omega^2=2.40$ to $\omega^2=3.61$. The overall shape and character of the mobility edge trajectory still remain largely unchanged in the finite inter-branch coupling case. In particular, the re-entrant behavior of the mobility edge around $V=0.5$
persists at these coupling parameters. Overall, our results suggest that coupling between different vibrational branches does not significantly alter the very nature of the Anderson transition in phonons with mass disorder for the box disorder with re-entrance behavior being present in both coupled and decoupled cases.
%This conclusion is further supported by our analysis of the binary disorder case discussed in Section $C$, where inter-branch coupling similarly had minimal impact on phonon localization.

\section{Conclusions}
\label{sec: conclusion}

In our study, we have introduced the multi-branch DCA and TMDCA formalisms to investigate the effects of disorder on the phonon spectrum in multi-branch systems. By comparing our results with exact ED results, we have found excellent agreement between the developed multi-branch DCA formalism and spectral functions for both binary and box disorder distributions. We have demonstrated the importance of the cluster extension of a single-site local CPA approximation to avoid averaging out essential non-local features in the spectrum. Although cluster DCA has advantages in incorporating non-local spatial correlations, it fails to capture Anderson localization due to the arithmetic averaging over disorder configurations. To address this limitation, we have developed the TMDCA multi-branch formalism, where a typical averaging ansatz replaces the arithmetic averaging in the self-consistency loop. Using the typical density of states as the order parameter for Anderson localization, we have investigated the impact of various vibrational multi-branch couplings ($\tau$ and $\varepsilon$) on phonon localization. Our study reveals that the multi-branch couplings in the model investigated have little effect on the Anderson transition of phonon localization in the presence of mass disorder. Specifically, our results demonstrate that localized states remain localized and extended states remain extended, largely regardless of the value of $\tau$ and $\varepsilon$. 
We also show that, for the box disorder distribution, inter-branch coupling does not alter the re-entrant behavior of the mobility edge, preserving the overall nature of the Anderson transition observed in the decoupled case. These results do not rule out that multi-branch couplings in other models could be important for the localization of phonons. We leave, that question for future studies. 

%\parskip
%\vspace{-0.39in}
%\quad 

In summary, our study demonstrates that the developed multi-branch DCA and TMDCA methods are computationally efficient, able to capture exact results even at a small cluster size of $N_c=4^3$, and are fully causal. The DCA is effective in capturing non-local spatial correlations and other fine features in the DOS, while the TMDCA can address the key aspects of the Anderson localization of phonons caused by the complex force-constants model. Therefore, the developed multi-branch cluster approach is a promising tool for investigating phonon localization in real materials. Furthermore, as Green's function approach, the multi-branch TMDCA can be readily applied to a range of layered geometries, superlattice structures, heterostructures, thin films, and interfaces.

\section*{acknowledgments} 
WM acknowledge support by NSF DMR 1944974 grant (method development). HT and TB acknowledge support by the U.S. Department of Energy, Office of Science, under award number DE-SC0025748 grant.
%WM acknowledges support by NSF OAC-1931367 grant.
This work used Expanse at SDSC through allocation DMR 130036 from the Advanced Cyberinfrastructure Coordination Ecosystem: Services \& Support (ACCESS) program, which is supported by National Science Foundation grants \#2138259, \#2138286, \#2138307, \#2137603, and \#2138296. This research also used resources of the Oak Ridge Leadership Computing Facility, which is a DOE Office of Science User Facility supported under Contract DE-AC05-00OR22725. A portion of this research (TB) was conducted at the Center for Nanophase Materials Sciences, which is a Department of Energy (DOE) Office of Science User Facility. In addition, we used resources of the National Energy Research Scientific Computing Center, a DOE Office of Science User Facility supported by the Office of Science of the U.S. DOE under Contract No. DE-AC02-05CH11231.

%\end{acknowledgments}

\appendix

\begin{figure}[!h]
\includegraphics[width=0.4\textwidth]
{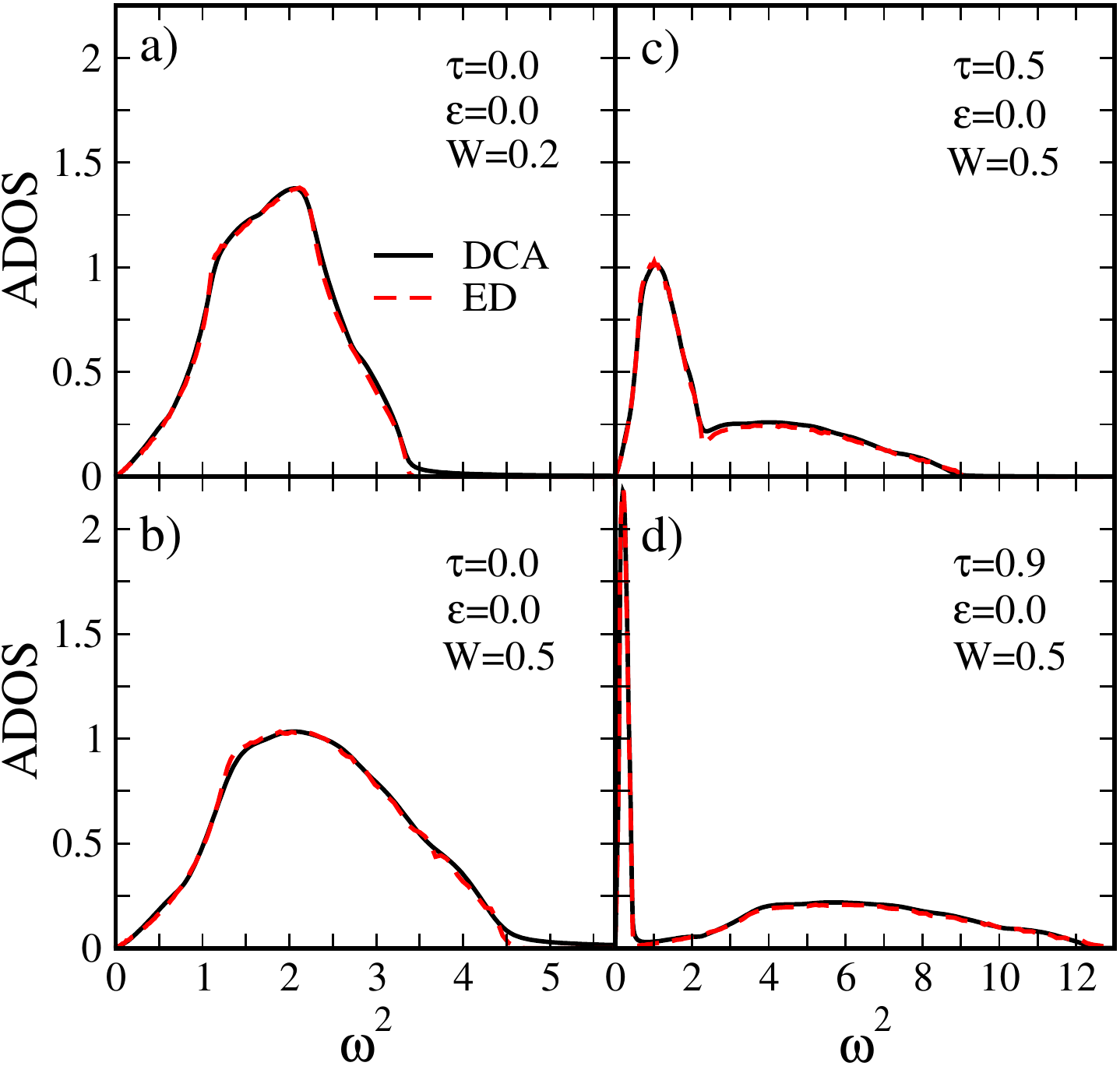}
\caption{
A comparison of the ADOS obtained from the multi-branch DCA using a cluster size of $N_c=4^3$ with the exact diagonalization method (ED) for the uniform (box) disorder distribution.
}
\label{fig:box_bench}
\end{figure}

\section{Comparison of the DCA and ED for the uniform (box) disorder distribution}
%Comparison of the ADOS obtained from the multi-branch DCA with exact results for uniform disorder distribution
%}

%\begin{figure}[!t]
%\includegraphics[width=0.4750\textwidth]
%{Fig7clusterconvergeold.eps}
%\caption{
%Parameters: $c=0.5$, $V=0.9$, $\tau=0.3$ and $\epsilon$=0.4 The evolution of TDOS with increasing cluster size N$_c=1, 4^3, 5^3$.}
%\label{fig:TDOSclusterconv}
%\end{figure}

To demonstrate the generalizability of our multi-branch DCA method, we compared its performance with exact diagonalzaition (ED) results for a uniform "box" disorder distribution ($P(V) = \frac{1}{W} \Theta(|V - W/2|) $). The left panel of Fig.~\ref{fig:box_bench} (a-b) shows the comparison for the case of $\tau=0$, whereas the right panel  (c-d) shows the comparison for finite values of $\tau=0.5, 0.9$. In both cases, we find excellent agreement with the exact ED results, indicating the validity of our multi-branch DCA method for arbitrary values of $\tau$ and strength of disorder $W$ in the case of uniform disorder distribution.

\section{Cluster-size convergence of the TDOS}

\begin{figure}[!th]
 %\vspace{-0.4in}
\includegraphics[width=0.4\textwidth] {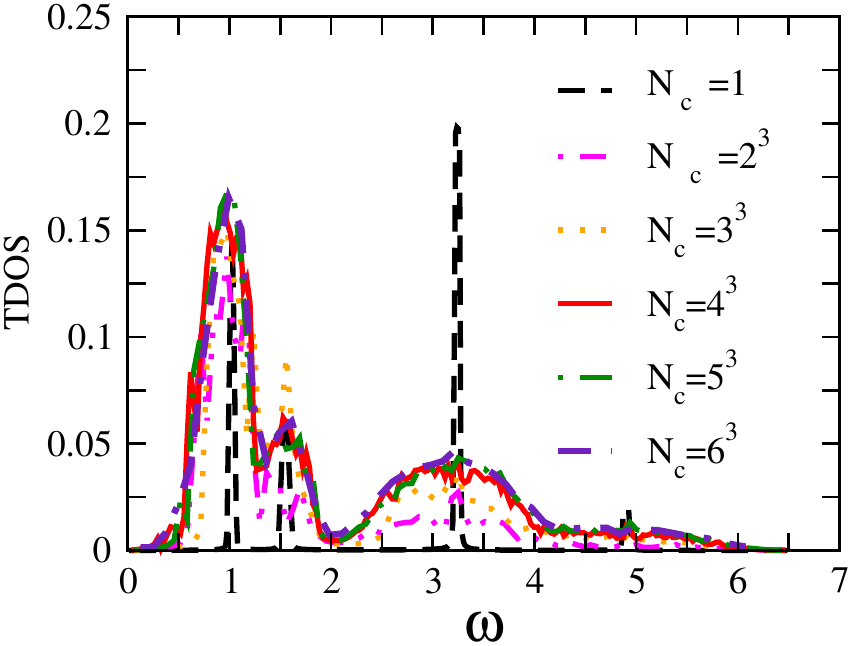}
\caption{Parameters: $c=0.5$, $V=0.9$, $\tau=0.3$ and $\epsilon$=0.4 The evolution of TDOS with increasing cluster size N$_c=1^3, 2^3,3^3,4^3,5^3$ and $6^3$.}
%\vspace{-0.4in}
\label{fig:TDOSclusterconv}
\end{figure}

%To demonstrate the cluster size convergence of our results, here, as an example, in Fig.\ref{fig:TDOSclusterconv}, we show the convergence of TMDCA results with increasing 
%$N_c$ in binary disorder ($c=0.5,V=0.9$) in the presence of finite coupling($\tau=0.3, \varepsilon=0.4$). As seen from Fig.\ref{fig:TDOSclusterconv}, isolated impurity modes appear in the high-frequency region ($2<\omega<5$) under strong disorder. The TDOS for 
%for N$_c=1$ almost vanishes at these frequencies (e.g., $2<\omega<3$, $3<\omega<4.8$, $5<\omega<6$), indicating strong localization of the corresponding phonon modes. However, as the cluster size N$_c$ increases, the TDOS starts to become finite at these frequencies indicating delocalized states. \textcolor{black} {In addition, we observe no significant changes in the TDOS for cluster sizes of $N_c = 4^3$ and larger, confirming the convergence of our calculations with increasing $N_c$.}
\textcolor{black} {
To illustrate the cluster-size convergence of our results for the TDOS, Fig.~\ref{fig:TDOSclusterconv} shows the evolution of the spectra with increasing $N_c$ for a representative binary disorder case ($c = 0.5$, $V = 0.9$) in the presence of finite coupling ($\tau = 0.3$, $\varepsilon = 0.4$). As seen in the figure, isolated impurity modes appear in the high-frequency range ($2 \lesssim \omega \lesssim 5$) under strong disorder. For $N_c = 1$, the TDOS is nearly zero throughout these intervals (e.g., $2 < \omega < 3$, $3 < \omega < 4.8$, $5 < \omega < 6$), reflecting strong localization of the corresponding phonon modes. As the cluster size increases, the TDOS acquires finite spectral weight in these same regions, signaling the emergence of extended states. For $N_c \ge 4^3$, the TDOS exhibits no further significant changes, demonstrating convergence with respect to cluster size.
}

%\vspace{-0.4in}

\section{ $\varepsilon=0$ and $\tau\neq 0$: branch-decoupled case}

{\bf Average medium case:} Suppose that in the $N_b$-branch force constant matrix, the branch and lattice degrees of freedom are decoupled as follows:

\begin{equation}
\Phi_{N_b}^{\alpha \beta}(l,l') = K_1(l,l')M(\alpha,\beta),
\end{equation}
here, $l$ and $\alpha$ denote the lattice and branch indices, respectively. $K_1(l,l')$ represents a single-branch force constant matrix in lattice-space, while $M(\alpha,\beta)$ is a $N_b \times N_b$ force constant matrix in the branch space, the latter with eigenvalues $\varepsilon_m$. 
In this case, we can show that the total multi-branch density of states, denoted as ADOS($\omega$), is related to the single-branch density of states, ados($\omega$), and the eigenvalues $\varepsilon_m$, as follows:
\begin{equation}
\text{ADOS}(\omega)= \sum_m \frac{1}{\sqrt{\varepsilon_m}} \text{ados}(\frac{\omega}{\sqrt{\varepsilon_m}})
\label{eq:B2}
\end{equation}

Below, we provide the derivation of Eq.~\ref{eq:B2}. The eigenvalues $E_J$ of $\Phi^{\alpha \beta}_{N_b}$ are related to the eigenvalues $\varepsilon_j$ of the $K_1$ matrix and $\varepsilon_m$ of the $M$ matrix as $E_J = \varepsilon_m \varepsilon_j$. The corresponding eigenvectors of $\Phi^{\alpha \beta}_{N_b}(l,l')$ are given by $\phi_J(l,\alpha) = \psi_j(l)\gamma_m(\alpha)$, where $\psi_j(l)$ is the eigenvector of the single-branch $K_1$ matrix, and $\gamma_m(\alpha)$ is the eigenvector of the force constant matrix $M$ in branch space.

From here, we observe that the multi-branch density of states $DOS(\omega,l)$ at site $l$ is given by: 
\begin{align}
& \text{DOS}(\omega,l)  = \sum_{\alpha,J} \phi_J(l,\alpha) \phi^{\star}_J(l,\alpha) \delta(\omega - \sqrt{E_J}) \nonumber \\
&  = \sum_{\alpha, m, j} \psi_j(l)\gamma_m(\alpha) \psi^{\star}_j(l)\gamma^{\star}_m(\alpha) \delta(\omega -\sqrt{\varepsilon_m \varepsilon_j}) \nonumber \\
& = \sum_{m,j}\psi_j(l) \psi^{\star}_j(l) \frac{1}{\sqrt{\varepsilon_m}}\delta(\frac{\omega}{\sqrt{\varepsilon_m}} -\sqrt{\varepsilon_j}) \nonumber \\
& = \sum_{m} \frac{1}{\sqrt{\varepsilon_m}} \text{dos}(\frac{\omega}{\sqrt{\varepsilon_m}},l)
\label{eq: B3}
\end{align}
where we used the completeness of the branch space with $\sum_{\alpha} \gamma^{\star}_m(\alpha) \gamma_m(\alpha)$=1, and that $\text{dos}(\frac{\omega}{\sqrt{\varepsilon_m}},l)=\sum_{j}\psi_j(l) \psi^{\star}_j(l)\delta(\frac{\omega}{\sqrt{\varepsilon_m}} -\sqrt{\varepsilon_j})$ is the single-branch local density of states at site $l$.

Hence, we can deduce that the local multi-branch density of states, $\text{ADOS}(\omega)$, can be written as
\begin{align}
\text{ADOS}(\omega) & =\frac{1}{N_l} \sum_l \text{DOS}(\omega,l) \nonumber  \\
& = \sum_m \frac{1}{\sqrt{\varepsilon_m}} \big ( \frac{1}{N_l}\sum_l \text{dos}(\frac{\omega}{\sqrt{\epsilon_m}},l) \big) \nonumber 
  \\
& =\sum_m  \text{ados}(\frac{\omega}{\sqrt{\epsilon_m}}), 
\label{eq:B4}
\end{align}
where $\text{ados}(\frac{\omega}{\sqrt{\epsilon_m}})=\frac{1}{\sqrt{\varepsilon_m}}\big ( \frac{1}{N_l}\sum_l \text{dos}(\frac{\omega}{\sqrt{\epsilon_m}},l) \big) $ is the average density of
states of the single branch.

\vspace{0.1in}
{\bf Typical medium case:} In the following, we show that, in general, there is no simple relation for the total typical density of states case for multi-branch systems, unlike the relation (Eq.~\ref{eq:B4}) for the average density of states case. To illustrate this, let's consider a case with  two lattice sites and two branches. Using the geometric averaging to calculate the typical DOS, with $\text{TDOS}(\omega)=\big ( \prod_l\text{DOS}(\omega,l)\big )^{\frac{1}{N_l}}$, we obtain  
$\text{TDOS}(\omega)  = \big ( {\text{DOS}(\omega,l_1) \text{DOS}(\omega,l_2) } \big )^{1/2}.$
Substituting in the expression for the $\text{DOS}(\omega,l)$ from Eq.~\ref{eq: B3}, we  have
\begin{align*}   
&\text{TDOS}(\omega)  = \lbrack {\text{DOS}(\omega,l_1) \text{DOS}(\omega,l_2) } \rbrack ^{1/2} \nonumber \\
& = \big \lbrack \big( \frac{1}{\sqrt{\varepsilon_1}}\text{dos}(\frac{\omega}{\sqrt{\varepsilon_1}},l_1) + \frac{1}{\sqrt{\varepsilon_2}}\text{dos}(\frac{\omega}{\sqrt{\varepsilon_2}},l_1) \big )\nonumber \\
& \times \big( \frac{1}{\sqrt{\varepsilon_1}}\text{dos}(\frac{\omega}{\sqrt{\varepsilon_1}},l_2) + \frac{1}{\sqrt{\varepsilon_2}}\text{dos}(\frac{\omega}{\sqrt{\varepsilon_2}},l_2) \big ) \big \rbrack^{1/2} 
\nonumber \\ 
& = \big \lbrack  \frac{1}{\sqrt{\varepsilon_1}}  \frac{1}{\sqrt{\varepsilon_1}}  \text{dos}(\frac{\omega}{\sqrt{\varepsilon_1}},l_1)\text{dos}(\frac{\omega}{\sqrt{\varepsilon_1}},l_2)\nonumber \\
& + \frac{1}{\sqrt{\varepsilon_1}}  \frac{1}{\sqrt{\varepsilon_2}} \text{dos}(\frac{\omega}{\sqrt{\varepsilon_1}},l_1)\text{dos}(\frac{\omega}{\sqrt{\varepsilon_2}},l_2)\nonumber \\
& + \frac{1}{\sqrt{\varepsilon_2}}  \frac{1}{\sqrt{\varepsilon_1}} \text{dos}(\frac{\omega}{\sqrt{\varepsilon_2}},l_1)\text{dos}(\frac{\omega}{\sqrt{\varepsilon_1}},l_2)\nonumber \\
\end{align*}

\begin{align}
%&\text{TDOS}(\omega)  = \lbrack {\text{DOS}(\omega,l_1) \text{DOS}(\omega,l_2) } \rbrack ^{1/2} \nonumber \\
%& = \big \lbrack \big( \frac{1}{\sqrt{\varepsilon_1}}\text{dos}(\frac{\omega}{\sqrt{\varepsilon_1}},l_1) + \frac{1}{\sqrt{\varepsilon_2}}\text{dos}(\frac{\omega}{\sqrt{\varepsilon_2}},l_1) \big )\nonumber \\
%& \times \big( \frac{1}{\sqrt{\varepsilon_1}}\text{dos}(\frac{\omega}{\sqrt{\varepsilon_1}},l_2) + \frac{1}{\sqrt{\varepsilon_2}}\text{dos}(\frac{\omega}{\sqrt{\varepsilon_2}},l_2) \big ) \big \rbrack^{1/2} 
\nonumber \\ 
%& = \big \lbrack  \frac{1}{\sqrt{\varepsilon_1}}  \frac{1}{\sqrt{\varepsilon_1}}  \text{dos}(\frac{\omega}{\sqrt{\varepsilon_1}},l_1)\text{dos}(\frac{\omega}{\sqrt{\varepsilon_1}},l_2)\nonumber \\
%& + \frac{1}{\sqrt{\varepsilon_1}}  \frac{1}{\sqrt{\varepsilon_2}} \text{dos}(\frac{\omega}{\sqrt{\varepsilon_1}},l_1)\text{dos}(\frac{\omega}{\sqrt{\varepsilon_2}},l_2)\nonumber \\
%& + \frac{1}{\sqrt{\varepsilon_2}}  \frac{1}{\sqrt{\varepsilon_1}} \text{dos}(\frac{\omega}{\sqrt{\varepsilon_2}},l_1)\text{dos}(\frac{\omega}{\sqrt{\varepsilon_1}},l_2)\nonumber \\
& + \frac{1}{\sqrt{\varepsilon_2}}  \frac{1}{\sqrt{\varepsilon_2}} \text{dos}(\frac{\omega}{\sqrt{\varepsilon_2}},l_1)\text{dos}(\frac{\omega}{\sqrt{\varepsilon_2}},l_2) \big \rbrack ^{1/2} \nonumber \\
%& = \prod_m (1/\sqrt(\varepsilon_m))\text{tdos}(\omega/\sqrt(\varepsilon_\alpha)) + (\text{cross terms} ) .
\label{eq: B6}
\end{align}
Now, considering that the typical density of states of an individual branch is defined as 
\begin{equation}\text{tdos}(\omega/\sqrt{\varepsilon_m})=\big ( \prod_l \frac{1}{\sqrt{\varepsilon_m}}\text{dos}(\omega/\sqrt{\varepsilon_m},l) \big )^{1/N_l}, \end{equation} we can rewrite Eq.~\ref{eq: B6} as
\begin{align}
&\text{TDOS}(\omega)  
%& = \sum_m \text{tdos}(\omega/\sqrt{\varepsilon_m} )
%+ \text{cross terms}.
& = \left( \sum_m \text{tdos}^2(\omega/\sqrt{\varepsilon_m} )
+ \text{cross terms}\right)^{1/2}.
%(sum_m tdos^2(w/sqrt(eps_m))+ cross-terms)^1/2
\label{eq: B7wrong}
\end{align}

%So, in addition to the first term, that is analogous with the right-hand side of Eq.~\eqref{eq:B4}, there will, in general, be cross-terms. Note, that this equation holds for the $N_b$ multi-branch case with N$_l$ sites. 
This in general is not equal to the typical density of states analogy of Eq.~\ref{eq:B4}:
\begin{align}
\text{TDOS}(\omega) = \sum_m \text{tdos}(\omega/\sqrt{\varepsilon_m} )
\label{eq: B7}
\end{align}
with 
$\text{tdos}(\omega/\sqrt{\varepsilon_m})=\big ( \prod_l \frac{1}{\sqrt{\varepsilon_m}}\text{dos}(\omega/\sqrt{\varepsilon_m},l) \big )^{1/N_l} .$  Note, that this equation holds for the $N_b$ multi-branch case with N$_l$ sites.

%\begin{figure}[t!]
%\vspace{-0.6in.}
%\includegraphics[width=0.48\textwidth]{figure/Fig7-appendix-v2.eps}
%{figure/FigTDOSbench.png}
%{figure/Fig1_3.png}
%\caption{
%Parameters: $\varepsilon=0.0, c_A=0.2, V_A=0.9, N_c=4^3$. A comparison of the TDOS obtained from analytical (extrapolated) expression (Eq.~\ref{eq:tdos_branch_relation}) and from the multi-branch TMDCA formalism (computed) for several values of $\tau=0.0, 0.1, 0.2, 0.3$.
%}
%\label{fig:TDOS_bench}
%\vspace{-0.1in.}
%\end{figure}

\bibliography{main}

\end{document}